\documentclass[useAMS,usenatbib]{mn2e}

\usepackage{epsfig}
\title[Compact stellar systems and cluster environments]{Compact stellar systems and cluster environments}
\author[P. Firth and M. J. Drinkwater and A. M. Karick]{P. Firth$^{1}$\thanks{E-mail:
firth@physics.uq.edu.au (PF)}, M. J. Drinkwater$^{1}$ and A. M. Karick$^{2}$\\
$^{1}$Department of Physics, University of Queensland, QLD 4072, Australia\\
$^{2}$Physics Department, U.C. Davis and IGPP, Lawrence Livermore National Laboratory, L-413 Livermore, CA 94550, USA}
\begin{document}

\bibliographystyle{mn2e}

\date{Accepted 2008 June 27. Received 2008 June 12; in original form 2008 March 19}

\pagerange{\pageref{firstpage}--\pageref{lastpage}} \pubyear{2008}

\maketitle

\label{firstpage}

\begin{abstract}
We have extended the search for luminous ($M_{b_J} \le -10.5$) compact stellar systems (CSSs) in the Virgo and Fornax galaxy clusters by targeting with the recently commissioned AAOmega spectrograph three cluster environments -- the cluster cores around M87 and NGC 1399, intracluster space, and a major galaxy merger site (NGC 1316). We have significantly increased the number of redshift-confirmed CSSs in the Virgo cluster core and located three Virgo intracluster globular clusters (IGCs) at a large distance from M87 (154-173 arcmin or $\sim\!750$-850$\; \mbox{kpc}$) -- the first isolated IGCs to be redshift-confirmed. We estimate luminous CSS populations in each cluster environment, and compare their kinematic and photometric properties. We find that (1) the estimated luminous CSS population in the Virgo cluster core is half of that in Fornax, possibly reflecting the more relaxed dynamical status of the latter; (2) in both clusters the luminous CSS velocity dispersions are less than those of the cD galaxy GC system or cluster dE galaxies, suggesting luminous CSSs have less energetic orbits; (3) Fornax has a sub-population of cluster core luminous CSSs that are redder and presumably more metal-rich than those found in Virgo; (4) no luminous CSSs were found in a 10-20$\; \mbox{ arcmin}$ (60-130$\; \mbox{kpc}$) radial arc east of the 3 Gyr old NGC 1316 galaxy merger remnant or in the adjacent intracluster region, implying that any luminous CSSs created in the galaxy merger have not been widely dispersed.
\end{abstract}

\begin{keywords}
galaxies: clusters: individual: Fornax cluster, Virgo cluster -- galaxies: distances and redshifts -- galaxies: individual: M87, NGC 1399, NGC 1316 --.galaxies: kinematics and dynamics -- galaxies: star clusters.
\end{keywords}

\section{Introduction}
High-resolution simulations of structure formation in the accepted $\Lambda$CDM Universe\footnote{$H_0 = 73 \; \mbox{km} \, \mbox{s}^{-1}$, $\Omega_M = 0.27$, $\Omega_\Lambda = 0.73$} predict the present-day survival, around isolated large galaxies such as the Milky Way or M31, of many more satellite dark matter concentrations than are observed through their radiant baryons \citep*[see discussion by][]{Cote..2002}. Two mechanisms may explain this apparent `missing satellite' problem.

Firstly that the epoch of reionization \citep[see discussion by][]{Cote..2002} after the first stars formed at $z=12\pm2$ \citep{Moore..2006} halted the collapse of baryonic matter (hydrogen gas) into the smallest dark matter gravitational potential wells -- evidence supporting this hypothesis includes detection in the Local Group of dark matter dominated, very low surface brightness star clusters or dwarf galaxies \citep{Belokurov..2007}.

Secondly that in the dense environments of galaxy clusters small dark matter haloes, although surviving the gradual hierarchical galaxy merger process, have been disrupted through tidal interaction with the cluster core. The visible remnants of this destruction are field stars and compact stellar systems (CSSs), which comprise the globular clusters (GCs) and the more recently discovered massive but sub-galactic luminous CSSs. Simulations by \citet*{Henriques..2007} show that galaxy cluster field stars (the detectable intracluster light) from totally and partially disrupted dwarf galaxies can explain the observed shallow slope in the faint end of the cluster galaxy luminosity function. Tidal threshing simulations (\citealt*{Bekki..2001}; \citealt{Bekki..2003I}) predict the formation of luminous ($M_{b_J} \le -10.5$) CSSs through the partial disruption of nucleated dwarf galaxies, by stripping away the stellar envelope to leave the naked nucleus of an ultra-compact dwarf galaxy (UCD). This process may also account for the surviving GCs, so comparing the properties of luminous CSSs with their fainter GC counterparts may reveal whether they form a single population -- for example, in the Fornax Cluster core region the luminosity distribution of the most luminous CSSs seems to extend smoothly into the bright tail of the GC population (\citealt*{Mieske..2004I}; \citealt{Drinkwater..2004}).

While both these mechanisms may have played a role, the great distances to nearby galaxy clusters limit observable evidence to remnant intracluster light and CSSs. Since disruption efficiency depends on environmental factors such as density and cluster mass, we investigate the spatial distribution, kinematics and photometric properties of an increased sample of Virgo and Fornax redshift-confirmed luminous CSSs in three cluster environments.

\begin{itemize}
\item \textbf{Cluster Core Environment.}  In the nearby Universe CSSs are numerous in the dense environments surrounding giant elliptical galaxies at the cores of galaxy clusters. Disruption through tidal interaction with the deep gravitational potential well of these giant galaxies is greatly enhanced in the cluster core, so we expect to observe a greater density of remnant CSSs and intracluster light \citep{Bekki..2004}. The Virgo and Fornax clusters at distances less than $20 \; \mbox{Mpc}$ contain the central giant elliptical galaxies M87 and NGC 1399 that have been extensively imaged for point source CSS candidates. The rich Virgo Cluster containing several thousand galaxies is embedded in an extensive supercluster, whereas the Fornax Cluster containing several hundred galaxies is smaller and isolated. Fornax has approximately two times the central galaxy density but approximately half the velocity dispersion of Virgo -- this implies that Fornax is more dynamically relaxed than Virgo and we expect a more evolved CSS population. Observations in Virgo and Fornax provide an opportunity to compare CSS populations in two differing cluster core regions, since both clusters are at similar distances and close enough to explore the faint end of their CSS distributions.

The cluster core environment is dominated by the extensive GC system of the central elliptical galaxy. These GCs are thought to have multiple origins -- most were formed during the first mergers of galaxies that formed the cD galaxy, and the remainder were stripped from galaxies accreted by the central giant elliptical galaxy as it evolved \citep[see review of GC formation mechanisms by][]{Ashman..1998}. GCs observed now are the survivors from an even larger population exposed to tidal destruction as the cluster core evolved. The complex and crowded environment in the cluster core blurs the forensic evidence of luminous CSS origins -- their distribution, kinematics and chemistry are consistent with both the tidal threshing and the bright GC theories.

The radial extent of cluster core environments can be approximated through imaging surveys of the fall in number density of point sources, predominantly GCs of the central giant elliptical galaxy, to a `background' level at which the number density profile flattens out. The transition to a background number density is difficult to precisely locate due to the relatively smooth decline in point source density and lack of supporting point source redshift measurements. Several photometric studies trace the overdensity of point sources across the core regions of Virgo and Fornax \citep[e.g.][]{Jordan..2002, Hasegan..2005, Bassino..2006}. For the spatially isolated Fornax Cluster, \citet{Bassino..2006} used this method to define a radial limit of $45\pm5 \, \mbox{arcmin}$ for the core region centred on  NGC 1399. The Virgo Cluster is part of a more extended and complex dynamical structure, making it more difficult to define a core region boundary -- however, several previous wide-field photometric and dynamical studies of M87's GC system \citep[e.g.][]{Cote..2001, Tamura..2006} show that it extends to at least the $60 \; \mbox{arcmin}$ radius of our AAOmega field.

\item \textbf{Intracluster Environment.} The intracluster environment beyond the tidal influence of the dominant cluster galaxies is relatively sparsely populated, mainly by dwarf elliptical (dE) galaxies. \citet{West..1995} proposed that galaxy clusters have a population of intracluster globular clusters (IGCs) not gravitationally bound to the central giant elliptical or other cluster galaxies. Subsequent photometry in the central region of Fornax has suggested an excess of GC candidates near dwarf galaxies \citep{Bassino..2003} and between NGC 1399 and the nearby galaxy NGC 1387 \citep{Bassino..2006}, while in wide-field Virgo imaging by \citet{Tamura..2006} GC candidates were found at some distance from major galaxies. More recently \citet{Williams..2007a} detected four faint potential IGCs (not redshift confirmed) in HST images of the outer part of Virgo's cluster core region $\sim\!40 \; \mbox{arcmin}$ from M87. None of the existing CSS redshift surveys \citep[e.g.][]{Drinkwater..2000a, Mieske..2004I, Bergond..2007} target IGCs far outside the cluster core environment -- our AAOmega observations include both Virgo and Fornax intracluster fields starting at least $\sim350 \, \mbox{kpc}$ from the central giant elliptical galaxies.

\item \textbf{Galaxy Merger Environment.} The galaxy merger environment differs from the cluster core environment, being marked by the recent merging of massive gas-rich galaxies rather than ongoing accretion of many relatively small galaxies. When large galaxies interact or merge they potentially disrupt satellite objects, and gas-rich mergers such as the well-known Antennae galaxy pair undergo intense starburst activity along shock fronts. It has been predicted \citep{Kroupa..1998, Fellhauer..2002, Fellhauer..2005} that luminous CSSs can evolve from stellar superclusters created from the merging of young massive clusters (YMCs) formed in such galaxy mergers. 

NGC 1316, the radio source Fornax A in the Fornax Cluster, is a giant elliptical galaxy located over 1 Mpc or $3.7^\circ$ south-west of the cluster core \citep[see photometric study by][]{Schweizer..1980}. It shows evidence of a 3 Gyr old major merger event which resulted in the formation of  massive star clusters \citep{Goudfrooij..2001a, Goudfrooij..2001b}. Based on the number density profile of its GC population, NGC 1316's radial extent is estimated to be approximately $7 \; \mbox{arcmin}$ \citep[see figure 14 in][]{Goudfrooij..2001b}. In order to include more spatially dispersed merger products, we define the radial extent of the NGC 1316 galaxy merger environment to be $20 \; \mbox{arcmin}$ ($\simeq\!30 \; \mbox{kpc}$) encompassing the faint outer traces of its stellar envelope. With AAOmega we have undertaken a redshift survey of point sources within this radius to locate additional luminous CSSs produced by the galaxy merger.

The NW border of our M87 AAOmega field encompasses the strongly distorted galaxy NGC 4438 which may be undergoing a merger or high-speed interaction with the nearby galaxy NGC 4435 \citep[see for example][]{Panuzzo..2007}. We did not specifically target these galaxies as a merger site, but we note in passing that our AAOmega survey results show no overdensity of confirmed foreground or cluster point sources that might be associated with them.
\end{itemize}

In Section 2 we describe Virgo and Fornax observations during 2006 with the AAOmega multi-fibre spectrograph at the 3.9-m Anglo-Australian Telescope (AAT). The Virgo observations were completed in 2006, between March 28 and April 1. The scientific objectives were to (1) compare the distributions of bright and faint CSSs, in order to test the hypothesis that bright CSSs are simply the bright tail of the GC luminosity distribution \citep[e.g.][]{Mieske..2004I}; and (2) test the hypothesis that bright CSSs are the remnant nuclei of tidally stripped dE,N galaxies, and are consequently more widely distributed than classical GCs -- tidal threshing simulations by \citep{Bekki..2003I} predict a cut-off in the UCD distribution at $2.5^\circ$ from M87. Our Fornax observations during 2006, over 5 nights between December 12 and 16, were designed to (1) investigate the UCD--IGC interface by measuring the radial distribution and kinematics of CSSs around NGC 1399; and (2) locate CSSs in the intracluster and galaxy merger environments near NGC 1316.

In Section 3 we add our results to previous published data in order to investigate the spatial distribution, recession velocity dispersion and photometric properties of Virgo and Fornax CSSs in the defined cluster environments. Our findings are summarised in Section 4. Table \ref{table:parameters} lists the parameters for the Virgo and Fornax galaxy clusters that we assume throughout this paper.

\begin{table*}
\caption{Assumed Cluster Parameters}
\label{table:parameters}
\centering{
\small{
\begin{tabular*}{1.00\textwidth}
     {@{\extracolsep{\fill}}lcclcl}
\hline \hline
Cluster/Galaxy & Distance & \multicolumn{2}{c}{Distance Modulus} & \multicolumn{2}{c}{Recession Velocity} \\
 & (Mpc) & (mag) & & ($\mbox{km} \, \mbox{s}^{-1}$) & \\[3pt]
\hline \hline
Virgo (M87) & 16.7 & 30.97 & \citet{Mei..2007} & 1307$\pm$7 & \citet{Smith..2000}\\
Fornax (NGC 1399) & 18.3 & 31.35 & \citet{Richtler..2000} & 1425$\pm$4 & \citet{Graham..1998}\\
NGC 1316 & 22.8 & 31.79 & \citet{Richtler..2000} & 1760$\pm$10 & \citet{Longhetti..1998}\\[3pt]
\hline
\end{tabular*}
}}
\end{table*}

\section{AAOmega Observations and Redshift Measurements}
 
\subsection{Virgo CSS Target Selection}
\begin{figure}
\centering
 \includegraphics[width=8.5cm]{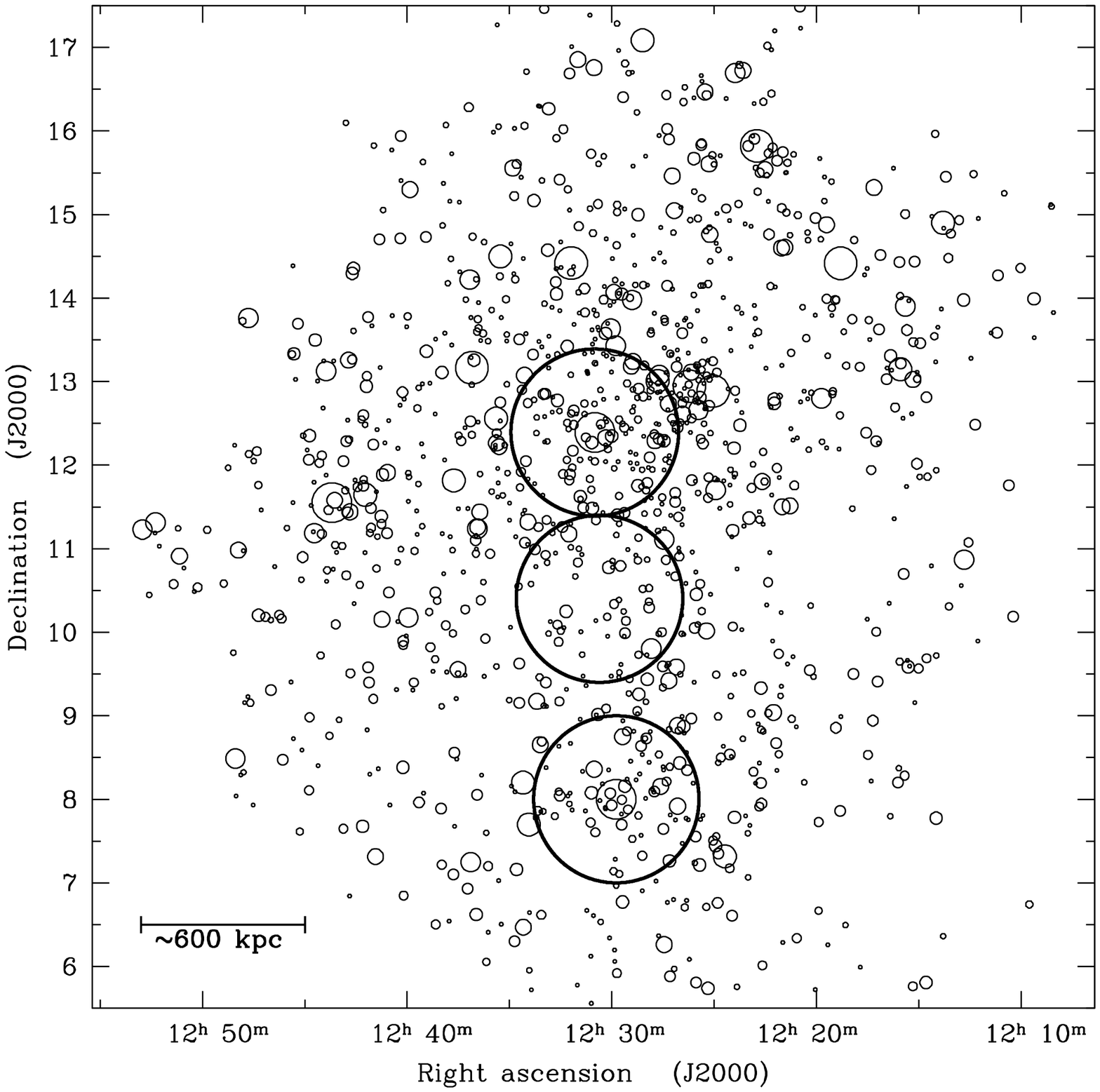}
 \includegraphics[width=8.5cm]{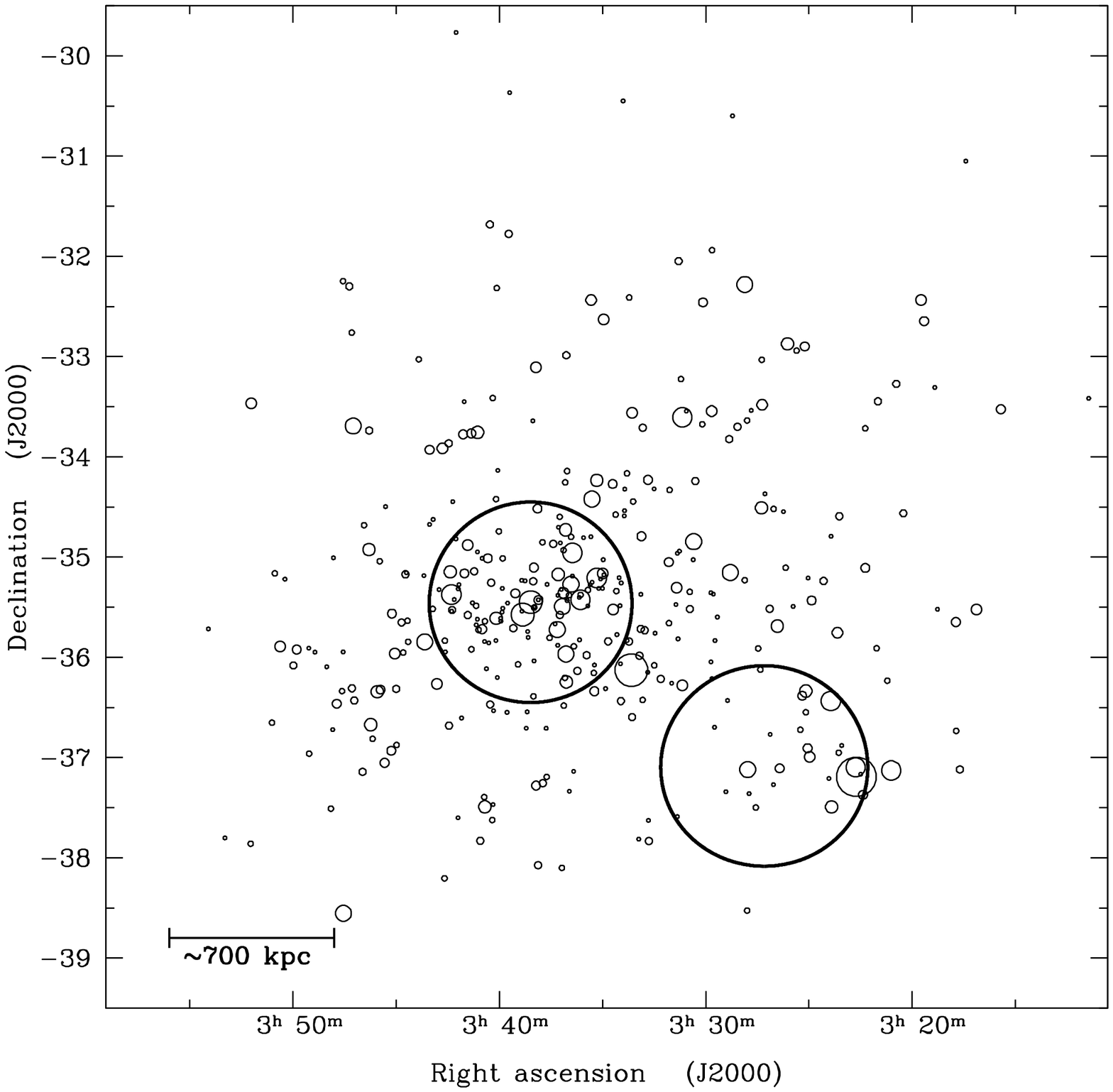}
 \caption{Prominent galaxies (circles with diameters indicating relative luminosity) in the regions covered by our proposed observations, scaled to appear at a common distance. \textsc{Upper:} From the Virgo Cluster Catalogue by \citet*{Binggeli..1985}, with three proposed AAOmega fields shown as $2^\circ$-diameter circles. The upper field is centred on M87 and the lower (unobserved) field on M49, with the intracluster field (12h30m) located between them. \textsc{Lower:} From the Fornax Cluster Catalogue by \citet{Ferguson..1989}, with the NGC 1399 cluster core field near the centre and the NGC 1316/intracluster field at lower right.}
 \label{fig:virgofields}
 \end{figure}

Fig.~\ref{fig:virgofields} (upper) shows the proposed AAOmega fields in their wider Virgo cluster context -- our point-source targets covered a wide area of the cluster between the massive elliptical galaxies M87 and M49, including the intracluster field 12h30m. We were unable to observe the M49 field due to adverse weather conditions.

We selected several thousand point source targets in each field using the same target selection method, described here for the M87 field as an example. From the Sloan Digital Sky Survey (SDSS DR4)\footnote{http://cas.sdss.org/dr4/en/} we obtained $gri$ magnitudes for 81159 objects in a $2^\circ$ diameter circular field centred on M87. We extracted a subset covering a $g$-band magnitude range of 17.2 to 21.8 ($B \simeq 17.0$ to 21.6), sufficiently faint ($M_B<-9.4$) to include both the luminous CSSs and the bright tail of the M87 GC luminosity distribution. At the Virgo Cluster distance most CSSs are unresolved point sources in ground-based telescopes. The SDSS classifies objects as stars (point sources) or galaxies (resolved objects) by comparing their magnitudes in $gri$ to magnitudes estimated from point-spread functions (psf) -- if the object's magnitude exceeds the psf-derived magnitude by more than 0.145 in a majority of pass-bands, it is classified as a galaxy. We relaxed this criterion to $\Delta\mbox{psf} > 0.225$ to ensure all likely CSSs would be targeted in the 12232 point-sources in the M87 field.

\begin{figure}
\centering
 \includegraphics[width=8.5cm]{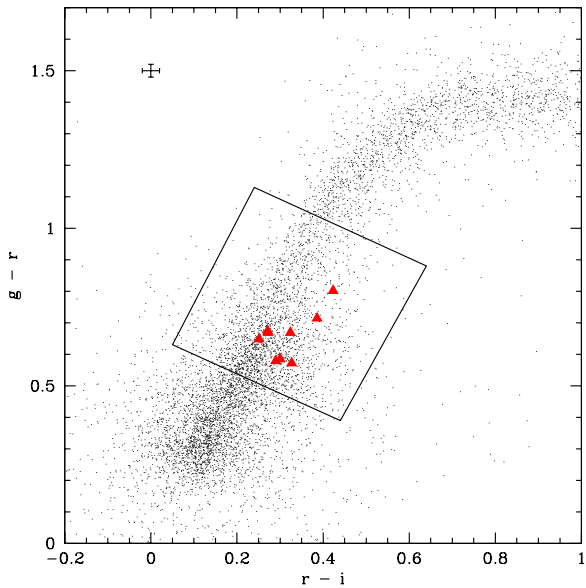}
 \caption{Colour distribution of SDSS DR4 point sources (dots) and previously known luminous CSSs \citep[triangles:][]{Jones..2006} surrounding M87 in the Virgo Cluster. The Galactic stellar locus slopes from lower-left to upper-right, where it turns horizontal due to the population of cool, red M-dwarf stars. The boxed area contains our potential targets, and the typical colour index error ($\sim0.02 \, \mbox{mag.}$) of all data points is shown at upper-left.}
 \label{fig:m87colorcolorplot}
 \end{figure}

Previous analysis in the Fornax Cluster \citep[PhD thesis by][]{Karick..2005} suggests that CSSs occupy a limited region of the $g - r$, $r - i$ colour plane.  Our Virgo Cluster colour selection box in Fig.~\ref{fig:m87colorcolorplot} (of the form $A \le gr + X(ri) \le B$, $C \le gr - Y(ri) \le D$ where $A,B,C,D,X,Y$ are suitably chosen constants) was designed to maximise coverage of faint CSS candidates around the colour distribution of known Virgo luminous CSSs. Redshift data were already available for many of these 2600 potential targets, leaving 2145 to be observed in this field. By this method we derived 2367 targets in the Virgo intracluster field and 3107 targets in the M49 field.

\subsection{Fornax CSS Target Selection} 
Fig.~\ref{fig:virgofields} (lower) shows our two Fornax AAOmega fields. Following our Virgo Cluster target selection approach, we used APM-derived\footnote{The Automatic Plate Measuring (APM) machine is a National Astronomy Facility run by the Institute of Astronomy in Cambridge. The APM Catalog can be found at http://www.ast.cam.ac.uk/$\sim\!$apmcat/.} $b_J$ and $R$ photometry from the southern-sky UK Schmidt Telescope photographic survey plates\footnote{For the UKST survey plates `blue' = $b_J$ (3900--5400\AA) and `red' = $R$ (5900--6900\AA) -- http://www.ast.cam.ac.uk/$\sim\!$mike/apmcat/apmcat$\_$params.html.}. Based on the colours of previously located Fornax CSSs, we selected 3780 point source CSS candidates in the NGC 1316 field and 2467 in the NGC 1399 field with $b_J - R < 1.7$. We restricted our targets to $b_J \le 20.8$ in the previously unsurveyed NGC 1316 field, but in the NGC 1399 field we extended the faint limit to $b_J \simeq 21.8$. Previous NGC 1399 2dF redshift surveys \citep{Drinkwater..2000b, Phillipps..2001} have greatly depleted the brighter point source targets compared with the virtually unobserved NGC 1316 field.

\subsection{Observing Conditions and Strategy}
Table \ref{table:aaoobs} summarises our completed observations in both clusters. Within each cluster all our targets were at essentially the same sky position, with airmass $<\!2$ for most of the night. We sorted potential targets into magnitude intervals and used the AAOmega signal-to-noise ratio (S/N) calculator to determine required exposure times under $2^{\prime\prime}$ seeing conditions in dark time. For the Virgo observations we aimed for S/N $\simeq\!8$ per $3.6 \; \mbox{\AA}$ resolution element (5.4 per $1.6 \; \mbox{\AA}$ pixel). This exceeds the minimum 2dF S/N of $\sim\!10$ per $9 \; \mbox{\AA}$ resolution element required for reliable redshift measurement using template cross-correlation \citep{Drinkwater..2000a}. To further improve redshift measurement success, we raised the S/N criterion for our subsequent Fornax AAOmega observations to 7 per $1.6 \; \mbox{\AA}$ pixel.

Using a beam splitter we obtained spectra from the blue (580V grating, $4799 \; \mbox{\AA}$ central wavelength) and red (385R grating, $7248 \; \mbox{\AA}$ central wavelength) AAOmega cameras. On each of two configuration plates, AAOmega provides 8 guide star fibres and was designed to provide 392 target fibres of $2^{\prime\prime}$ diameter over a $2^\circ$ field of view; however the available target fibres have been reduced by defects such as breakages and fringing effects. We allocated 25 fibres in each exposure to vacant sky positions identified from imaging, and configured the remaining available fibres on targets. Only a handful of fibres in each set were unallocated due to positioning constraints. With a minimum exposure time of 60 minutes to allow for fibre repositioning, and allowing for technical problems and weather delays, we expected to observe nearly all our intended targets. However, intermittent cloud and poor seeing reduced the number of observed targets and S/N of reduced spectra, forcing us to concentrate on the brighter end of our target range and to omit the Virgo M49 field. 

\begin{table*}
\caption{AAOmega Virgo and Fornax Cluster Observations}
\label{table:aaoobs}
\centering{
\small{
\begin{tabular*}{1.00\textwidth}
     {@{\extracolsep{\fill}}lccclc}
\hline \hline
Observation & UT Date & Targets & $B$-band Range & Exposure & Seeing$^a$\\
Set & 2006 &  & (mag) & (sec) & (arcsec)\\[3pt]
\hline \hline \\
\multicolumn{6}{l}{\textbf{Virgo M87 Field: 12:30:49.4, +12:23:28}}\\
set2-1 & 29 March & 328 & 17.0 -- 21.0 & $3\times1800$ & 1.8\\
set2-2 & 29 March & 328 & 17.0 -- 21.0 & $2\times1200$ & 1.8\\
set3-1 & 31 March & 329 & 17.0 -- 21.3 & $3\times1800$ & 2.0\\
set3-2 & 31 March & 329 & 17.0 -- 21.3 & $3\times1800$ & 4.0\\
\\
\multicolumn{6}{l}{\textbf{Virgo 12h30m Field: 12:30:00.0 +10:00:00}}\\
set2-1 & 1 April & 332 & 17.0 -- 20.1 & $2\times1800$ & 2.0\\
set2-1 & 1 April & 332 & 17.0 -- 20.1 & $1\times1800$ & 2.4\\
set3-1 & 1 April & 334 & 17.1 -- 20.7 & $3\times1800$ & 4.5\\
set4A-1 & 29 March & 343 & 17.5 -- 21.2 & \parbox[t]{3.5cm}{$3\times1800$, $1\times1000$} & 1.8\\
set4A-2 & 1 April & 343 & 17.5 -- 21.1 & $4\times1800$ & 2.4\\
set5-1 & 31 March & 341 & 17.2 -- 21.6 & $5\times1800$ & 2.0\\
set5-1 & 31 March & 341 & 17.2 -- 21.6 & $1\times1800$ & 4.0\\
\\
\hline\\
\multicolumn{6}{l}{\textbf{Fornax NGC 1399 Field: 03:38:29.1, -35:27:03}}\\
set1 & 12 December & 320 & 17.9 -- 21.5 & \parbox[t]{3.5cm}{$3\times2700$, $1\times1800$} & 4.0\\
set3 & 15 December & 319 & 16.3 -- 20.5 & $3\times720$ & 1.2\\
set3 & 16 December & 319 & 16.3 -- 20.5 & $1\times2400$ & 2.5\\
set4 & 15 December & 319 & 16.6 -- 21.4 & \parbox[t]{3.5cm}{$3\times2400$, $2\times1200$} & 3.0 -- 3.5\\
set4 & 16 December & 319 & 16.6 -- 21.4 & $4\times1500$ & 2.4\\
set5 & 15 December & 327 & 20.3 -- 21.5 & $4\times2400$ & 1.2 -- 2.6\\
set5 & 16 December & 327 & 20.3 -- 21.5 & \parbox[t]{3.5cm}{$2\times2700$, $3\times1800$} & 2.0\\
\\
\multicolumn{6}{l}{\textbf{Fornax NGC 1316 Field: 03:27:10.0 -37:05:00}}\\
set1 & 12 December & 328 & 16.3 -- 18.0 & \parbox[t]{3.5cm}{$2\times1200$, $1\times1500$} & 3.0\\
set2 & 12 December & 320 & 18.6 -- 19.5 & $3\times1800$ & 3.0\\
set3 & 13 December & 327 & 16.3 -- 18.3 & $4\times600$ & 3.0\\
set4 & 13 December & 319 & 18.8 -- 21.7 & $3\times1500$ & 3.0\\
set5 & 14 December & 328 & 17.4 -- 19.6 & \parbox[t]{3.5cm}{$5\times480$, $1\times260$} & 1.3\\
set6 & 14 December & 319 & 16.7 -- 20.4 & $5\times720$ & 1.3\\
\\
\hline
\end{tabular*}
\begin{flushleft}
\item[a.] We quote single figures for average seeing during each observation set, or a seeing range if it varied significantly between the exposures.
\end{flushleft}
}}
\end{table*}

\subsection{Signal-to-Noise Results}
Our Virgo observations have a small but systematic S/N variation across each field, possibly due to plate distortion or incorrect fibre placement in the AAOmega commissioning phase, which was corrected by the time of our Fornax observations. However, since many of our point source targets were background galaxies with strong emission lines, S/N was not the only factor affecting our ability to obtain redshift measurements. A 2D K-S test showed no significant difference (probability $<\!0.01$) in the distribution of point source targets across the Virgo intracluster field for which we obtained acceptable or unreliable redshift measurements.

\begin{figure*}
\centering
 \includegraphics[width=8.5cm]{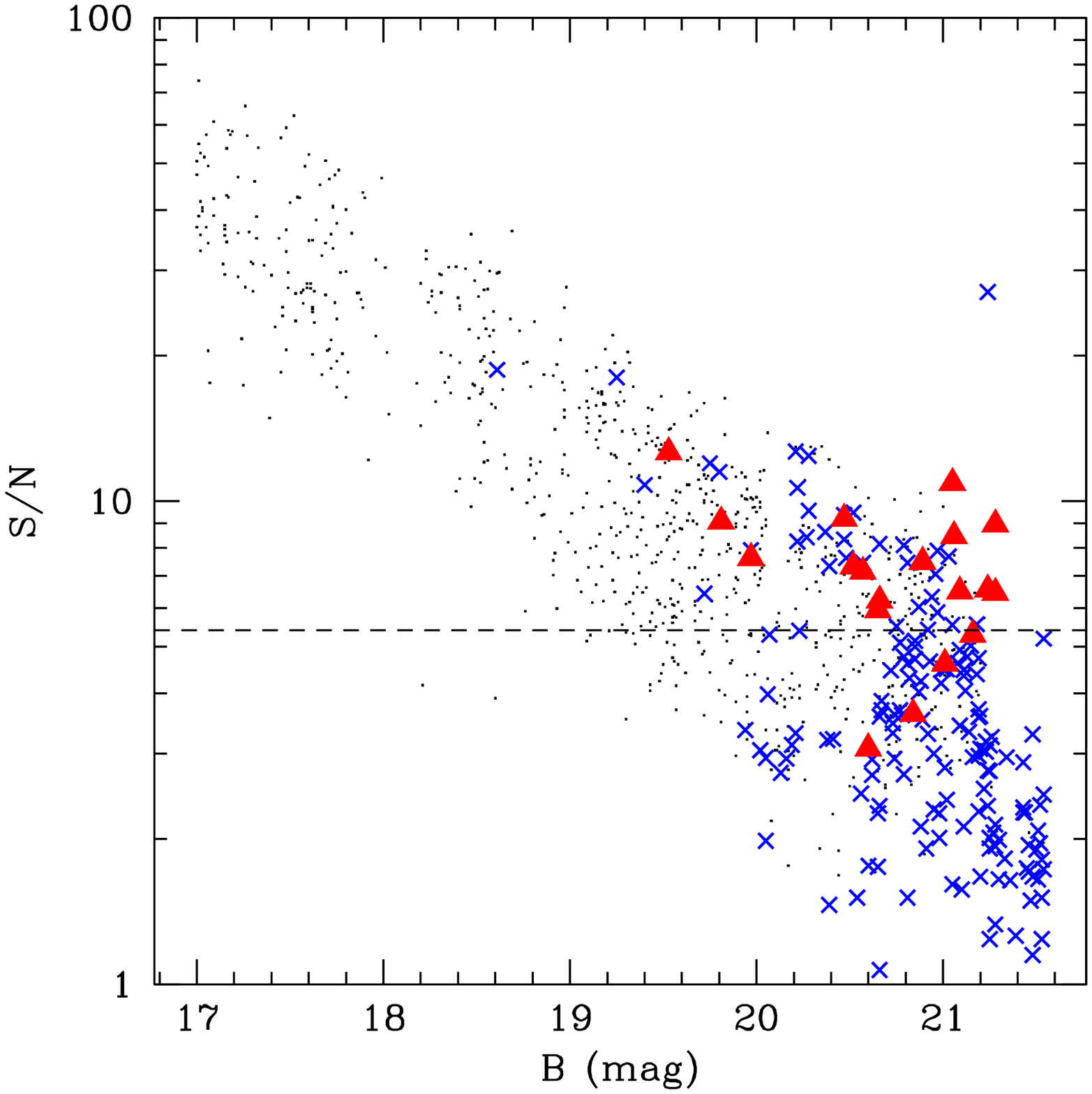}
 \includegraphics[width=8.5cm]{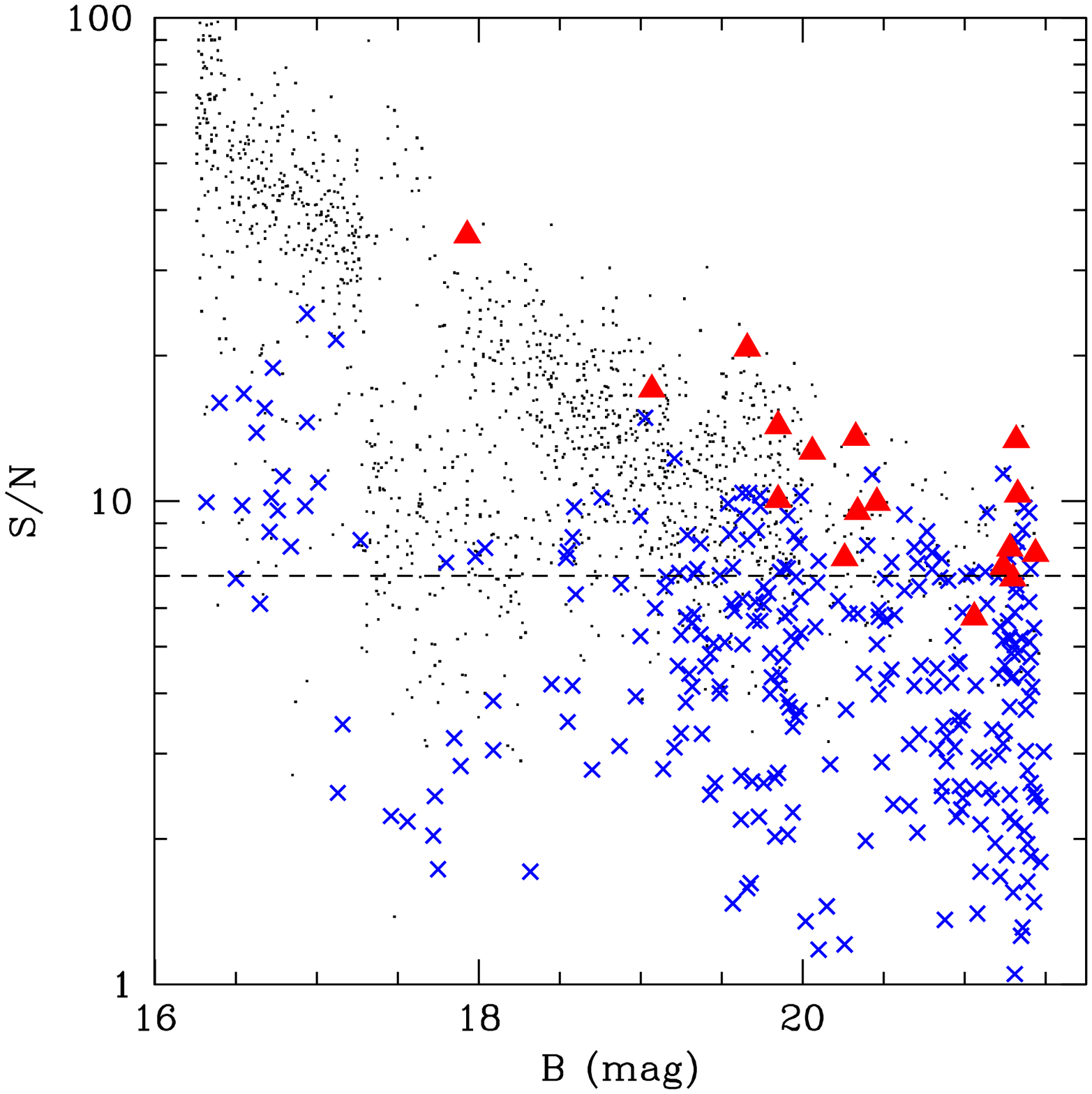}
 \vspace{0.5cm}
 \caption{S/N versus magnitude of spectra from our Virgo (left) and Fornax (right) AAOmega observations. Successfully measured redshifts from the spectra near or below our desired S/N minima (horizontal dashed lines) are biased strongly towards background QSOs and emission-line galaxies (crosses) with strong, easily identifiable emission features. Redshift-confirmed CSSs (filled triangles) are distributed mostly near the faint end of our target sets, with generally low but acceptable S/N. In the Fornax plot the drop in S/N at $B\simeq17.5$ is because NGC 1316 exposure set 3 was curtailed by cloud.}
 \label{fig:aaovirgofornaxsn2}
 \end{figure*}

The cluster core fields had fewer previously unredshifted bright point source targets than the intracluster fields, so the redshift measurement completeness of their predominantly faint targets should have been adversely affected by the observing conditions. However, a significant number of the low S/N spectra were background galaxies with strong emission line features from which we were able to successfully measure redshifts (see Fig.~\ref{fig:aaovirgofornaxsn2}). The results follow an expected trend of increasing S/N at brighter magnitudes, except for one abnormally high S/N data point in the M87 field which is part of a large spiral galaxy IC3489 (VCC 1490) mis-identified by SDSS as a point source (unlike point sources, for sufficiently extended objects no signal is lost from the fibre due to seeing). NGC 1316 exposure set 3 had uniformly low S/N because the exposure was curtailed by cloud. We obtained redshifts for some CSSs with sub-optimal S/N by using their Ca$\,$II triplet features.
 
\subsection{Data Reduction and Redshift Measurement}
Raw spectral images were reduced in a standard manner with \textsc{2dfdr}\footnote{The \textsc{2dfdr} software is available from ftp.aao.gov.au/pub/2df/.} and spliced to produce spectra which are continuous over the range 3800--$8900 \; \mbox{\AA}$. Initial redshift measurement was done in automatic mode using \textsc{runz} cross-correlation software\footnote{\textsc{RUNZ} is based on the cross-correlation method of \citet{Tonry..1979} and was developed by W.J. Saunders and others.} with standard star, emission-line galaxy and QSO templates. We then individually inspected each spectrum and corrected the redshift by identifying prominent absorption or emission lines. For QSOs we identified broad emission features (N$\,$V, Si$\,$IV, C$\,$IV, C$\,$III, Mg$\,$II, H$\beta$, O$\,$III) using a series of template spectra at increasing redshift. At low S/N where absorption features are difficult to detect, our successful redshift results are dominated by background galaxies with strong, easily identifiable emission features.

\begin{figure*}
\centering
\includegraphics[width=17cm]{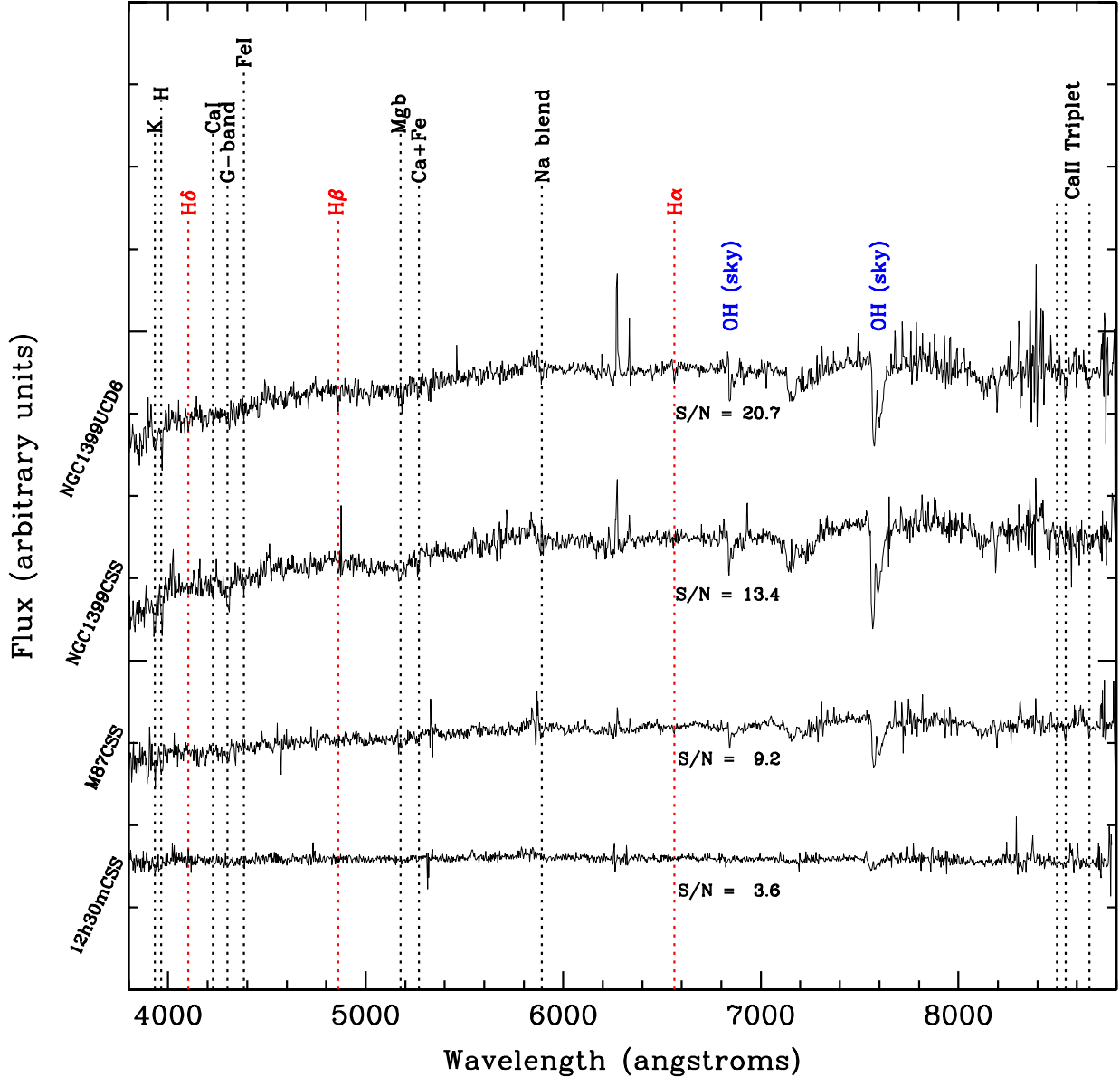}
\caption{Typical spectra of confirmed CSS (smoothed $\times 3$, de-redshifted to $cz=0$ and arbitrarily flux-scaled) at a range of indicated $\mbox{S/N} \, \mbox{pixel}^{-1}$ from the Fornax (NGC1399) and Virgo (M87) core and the Virgo intracluster (12h30m) environments. These can be found by their S/N values in the tables of CSS's detected (see later). Apparent emission lines in the UCD 6 spectrum around $6300 \; \mbox{\AA}$ and the large absorption feature near 6800 and $7600 \; \mbox{\AA}$ are residual sky features.}
\label{fig:targspectra}
\end{figure*}

Due to variable observing conditions and the magnitude range of each multi-fibre exposure set, we achieved a wide range of S/N in our target spectra. Fig.~\ref{fig:targspectra} compares the smoothed ($\times 3$) spectra of four successfully redshifted CSSs at various S/N levels. In general, our CSS spectra have insufficient S/N to make a detailed comparison of the spectral features between cluster environments.

\begin{figure*}
\centering
\includegraphics[width=8.5cm]{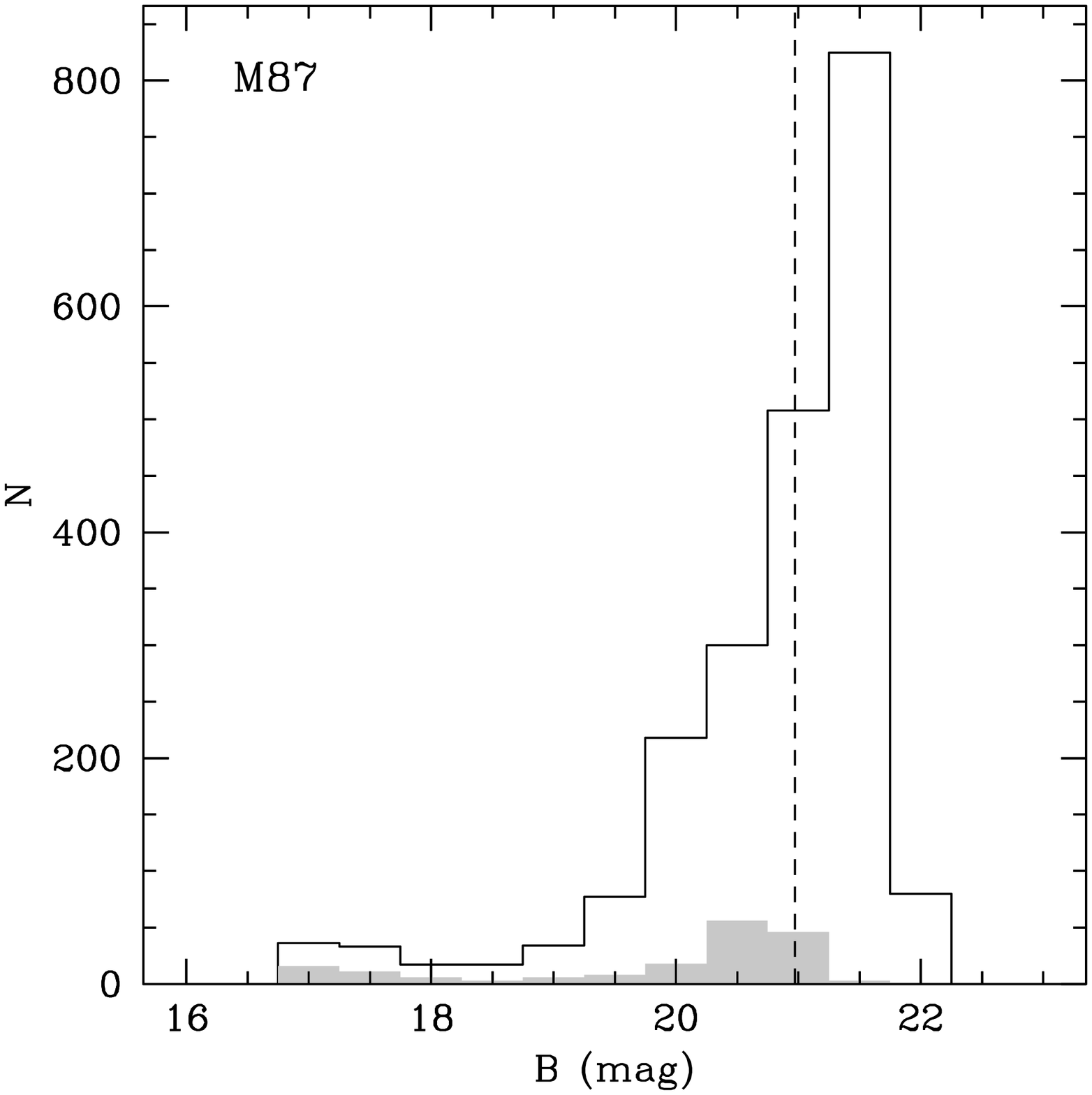}
\includegraphics[width=8.5cm]{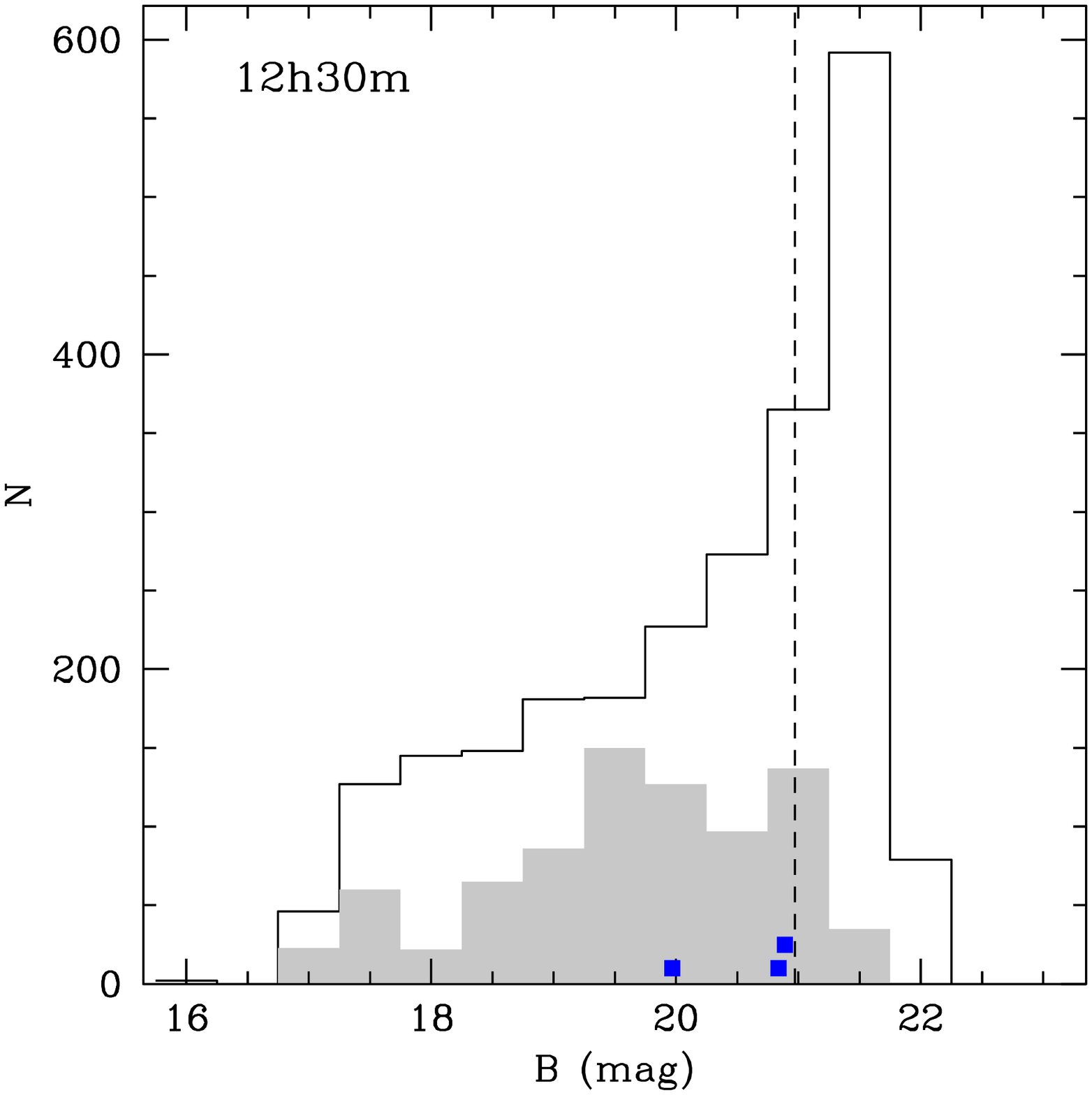}
\includegraphics[width=8.5cm]{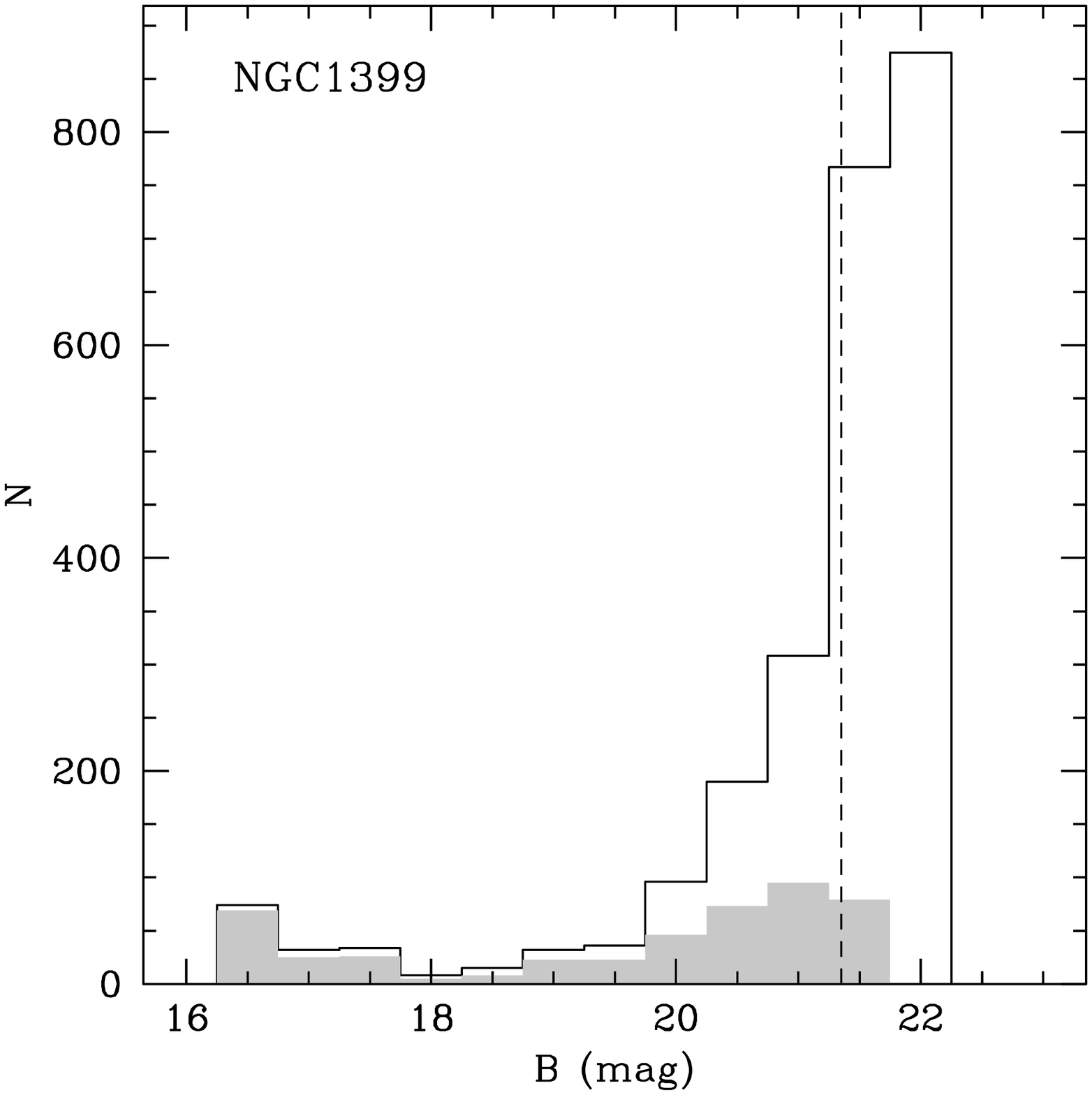}
\includegraphics[width=8.5cm]{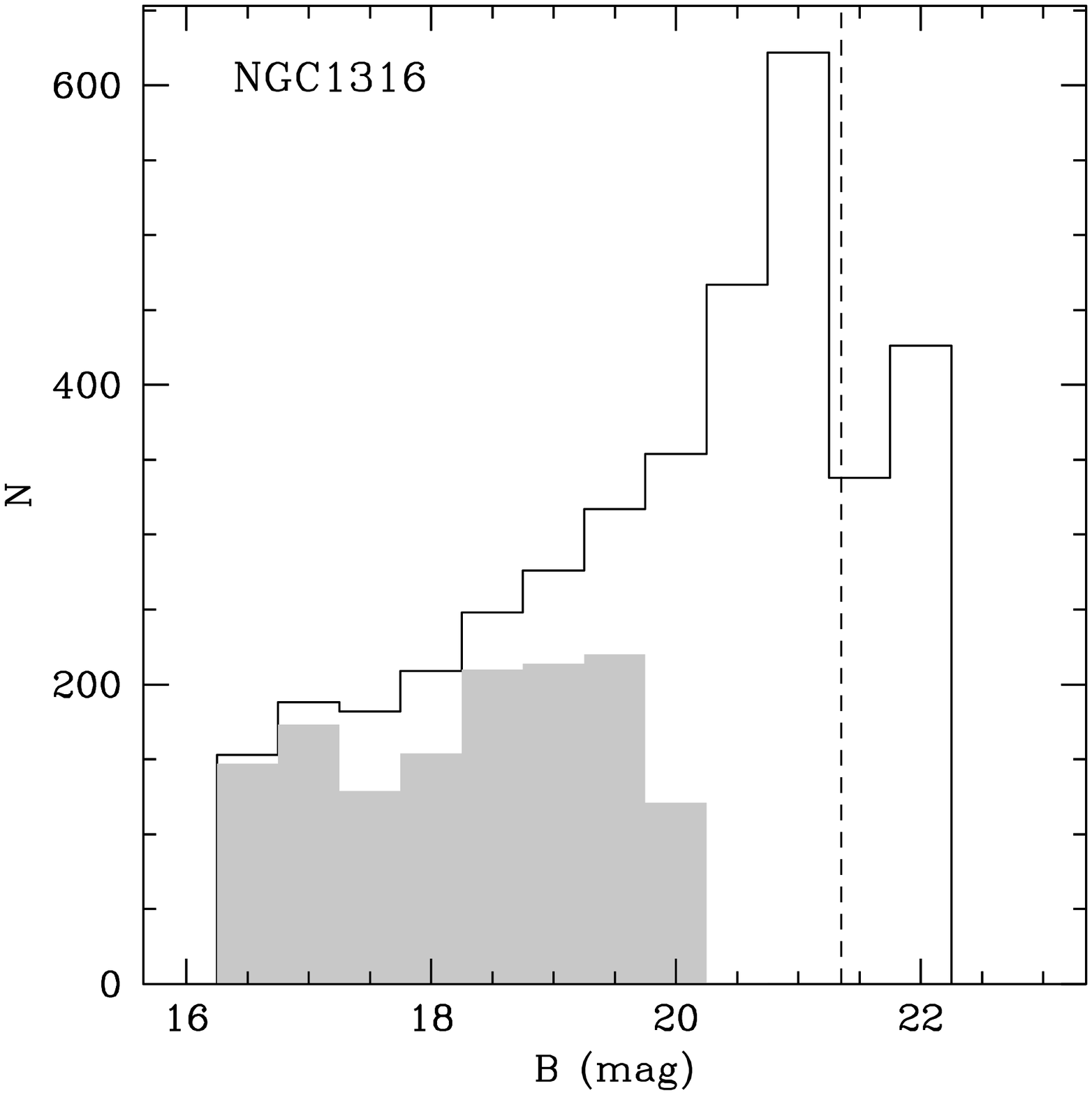}
\vspace{0.5cm}
\caption{Histograms of redshift completeness for our AAOmega observations, comparing potential targets (unshaded) with successful redshifts (shaded). The vertical dashed lines mark approximate boundary ($M_B=-10$) between GCs (to the right) and luminous CSSs (to the left). \textsc{Upper:} In Virgo, due to observing conditions, we concentrated on the previously unsurveyed intracluster field where we found three IGCs (squares). \textsc{Lower:} The brighter point sources in NGC 1399 have been covered in previous point-source redshift surveys, whereas the NGC 1316 field was previously unobserved. The bright redshifts in NGC 1399 include known CSSs which were being specifically reobserved to obtain high S/N spectra.}
\label{fig:aaovirgofornax_3}
\end{figure*}

The redshift measurement completeness of our observations varies between the four observed fields (see Fig.~\ref{fig:aaovirgofornax_3}). Both cluster core fields had already been extensively surveyed \citep{Drinkwater..2000a, Mieske..2004I, Jones..2006} leaving relatively few bright point source targets with unmeasured redshifts, whereas the two intracluster fields contained many bright point source targets with no measured redshift. Weather significantly limited our observing programs and reduced the intended redshift completeness.

To identify cluster members we define cluster recession velocity limits. The Fornax Cluster is isolated in redshift space from both Galactic stars and background galaxies, so with some confidence we define cluster membership by velocity limits of 500 to $2500 \; \mbox{km} \, \mbox{s}^{-1}$. The Virgo Cluster is embedded in the Virgo Supercluster which stretches to include the Local Group, and being less dynamically relaxed than Fornax it even includes galaxies with negative recession velocities. A few Virgo objects with `star-like' velocities may therefore be cluster CSSs, but considering the 500 to 2200$\; \mbox{km} \, \mbox{s}^{-1}$ velocity limits applied in the previous Virgo CSS survey \citep{Jones..2006}, we decided to adopt the same recession velocity limits as Fornax.

Having identified the cluster members, we rechecked their redshifts with \textsc{XCSAO}\footnote{\textsc{XCSAO} is a cross-correlation task available within \textsc{IRAF}.} cross-correlation software. In general, we achieved velocity errors $\le\!30 \; \mbox{km} \, \mbox{s}^{-1}$.

We examined SDSS or CTIO \citep*{Karick..2007} CCD images of each detected cluster object to confirm their compact nature and local environment. Two cluster objects in the Virgo intracluster field were parts of the spiral galaxy NGC 4451 which SDSS incorrectly identified as point sources. One new cluster object located in the Fornax NGC 1399 field was catalogued by APM but not identifiable on the CTIO image, and two other CSSs were outside the CTIO image areas. All other objects were compact, circular and featureless, except for one object in each Virgo field (see Table \ref{table:aaovirgolist}) which showed evidence of a possible outer stellar envelope.

\subsection{Results}
Table \ref{table:results} summarises our redshift measurement results and extinction-corrected photometry -- our selection criteria have successfully targeted samples of the CSS populations in both clusters. Tables \ref{table:aaovirgolist} and \ref{table:aaofornaxlist} provide details of the redshift-confirmed cluster objects. We found 18 new CSSs in the core regions of Virgo and Fornax, and 3 IGCs in the Virgo intracluster field. We found no CSSs in the Fornax intracluster/galaxy merger field, allowing an upper limit to be placed on their distribution. 
 
\begin{table*}
\caption{AAOmega Observing Results Summary}
\label{table:results}
\scriptsize{
\begin{tabular*}{1.00\textwidth}
     {@{\extracolsep{\fill}}llcccc|ccc}
\hline \hline
Field & Cluster & Potential & Targets & Successful & Success Rate & Galactic & Background & Cluster\\
Name & Environment & Targets & Observed$^a$ & Redshifts & (per cent) & Stars & Galaxies & Members\\[3pt]
\hline \hline
 & & & & & & & & \\
M87 & cluster core & 2145 & 657 & 173 & 26.3 & 120 & 37 & 16\\
12h30m & intracluster & 2367 & 1326 & 800 & 60.3 & 674 & 123 & 3\\
 & & & & & & & & \\
NGC 1399 & cluster core & 2467 & 1275 & 471 & 36.9 & 296 & 158 & 17\\
NGC 1316 & \parbox[t]{2cm}{intracluster} & 3564 & 1796 & 1304 & 72.6 & 1184 & 120 & 0\\
NGC 1316 & \parbox[t]{2cm}{galaxy merger} & 216 & 119 & 65 & 55.1 & 57 & 8 & 0\\
 & & & & & & & & \\
\hline
\end{tabular*}
\begin{flushleft}
\item a. Excluding duplicates of targets common to more than one observing set.
\end{flushleft}
}
\end{table*}

\begin{table*}
\caption{AAOmega Virgo Cluster CSSs Detected.}
\label{table:aaovirgolist}
\centering
\scriptsize{
\begin{tabular*}{1.00\textwidth}
     {@{\extracolsep{\fill}}llcccccr@{\extracolsep{0pt}}@{$\,\pm\,$}l@{\extracolsep{\fill}}cl}
\hline \hline
R.A.(J2000.0) & Dec.(J2000.0) & \multicolumn{5}{c}{SDSS Photometry} & \multicolumn{2}{c}{cz} & S/N & Notes\\
\cline{3-7}
(h :m :s ) & ($^\circ$ :$^\prime$ :$^{\prime\prime}$ ) & \textit{u} & \textit{g} & \textit{r} & \textit{i} & \textit{z} & \multicolumn{2}{c}{($\mbox{km} \; \mbox{s}^{-1}$)} & & \\[3pt]
\hline \hline\\
\multicolumn{6}{l}{\textbf{M87 Field}}\\
12:30:35.12 & 12:24:53.5 & 23.41 & 21.09 & 20.57 & 20.29 & 20.05 & 1636&22  & 6.5  & large velocity difference from M87$^a$\\ 
12:30:40.54 & 12:24:11.9 & 22.66 & 21.05 & 20.44 & 19.99 & 19.93 & 1391&36  & 10.9 & \\ 
12:30:41.20 & 12:19:46.2 & 23.02 & 21.24 & 20.56 & 20.06 & 19.72 & 1342&33  & 4.6  & \\ 
12:30:42.29 & 12:26:20.5 & 22.58 & 21.06 & 20.50 & 20.13 & 19.98 & 1506&27  & 8.5  & near galaxy 2MASX J12304203+1226263\\ 
12:30:46.78 & 12:15:32.1 & 21.80 & 20.52 & 19.91 & 19.63 & 19.51 & 1257&14  & 7.3  & \\ 
12:30:47.40 & 12:33:01.7 & 20.93 & 19.53 & 18.89 & 18.54 & 18.37 & 1379&23  & 12.6 & \\ 
12:30:52.79 & 12:16:00.8 & 22.63 & 21.16 & 20.63 & 20.43 & 20.15 & 1257&49  & 5.3  & \\ 
12:30:52.81 & 12:25:54.8 & 22.90 & 21.28 & 20.63 & 20.47 & 20.27 & 1113&27  & 9.0  & \\ 
12:30:57.48 & 12:31:39.5 & 22.79 & 20.65 & 20.02 & 19.73 & 19.64 & 882&33   & 6.0  & possible stellar envelope\\ 
12:30:58.08 & 12:20:20.2 & 22.59 & 21.28 & 20.72 & 20.59 & 20.42 & 1393&51  & 6.5  & \\ 
12:30:58.58 & 12:26:24.8 & 21.88 & 20.57 & 19.92 & 19.71 & 19.42 & 1984&35  & 7.1  & large velocity difference from M87\\ 
12:31:02.59 & 12:34:14.1 & 21.16 & 19.81 & 19.21 & 18.98 & 18.76 & 1192&18  & 9.1  & \\ 
12:31:05.13 & 12:20:03.5 & 22.37 & 20.47 & 19.75 & 19.33 & 19.11 & 1258&15  & 9.2  & DGTO (S314) 1220.6$\pm$10.1$^b$\\ 
12:31:05.80 & 12:26:40.7 & 21.70 & 20.66 & 19.97 & 19.64 & 19.40 & 901&17   & 6.2  & large velocity difference from M87\\ 
12:32:02.53 & 11:56:25.9 & 23.58 & 21.01 & 20.29 & 19.98 & 19.76 & 835&37   & 4.6  & large velocity difference from M87\\ 
12:33:29.47 & 12:10:17.5 & 22.86 & 20.60 & 20.00 & 19.76 & 19.68 & 1317&29  & 3.1  & \\ 
\\
\multicolumn{6}{l}{\textbf{Intracluster Field 12h30m +10d}}\\
12:29:50.62 & 09:31:03.7 & 22.01 & 20.89 & 20.38 & 20.05 & 19.89 & 1056&43  & 7.5  & \\ 
12:32:55.55 & 09:44:12.7 & 22.28 & 20.84 & 20.26 & 19.92 & 19.75 & 1743&33  & 3.6  & \\ 
12:33:07.36 & 09:52:54.3 & 21.21 & 19.97 & 19.39 & 19.01 & 18.86 & 1101&24  & 7.6  & possible stellar envelope; nearby red object\\ 
\\
\hline
\end{tabular*}
\begin{flushleft}
\item a. The recession velocity of the central cluster galaxy M87 is $1307\pm7$ \citep{Smith..2000}.\\
\item b. This object was previously detected \citep{Hasegan..2005} in the ACS Virgo Cluster survey.
\end{flushleft}
 }
\end{table*}

\begin{table*}
\caption{AAOmega Fornax Cluster CSSs Detected}
\label{table:aaofornaxlist}
\centering
\scriptsize{
\begin{tabular*}{1.00\textwidth}
     {@{\extracolsep{\fill}}llccccr@{\extracolsep{0pt}}@{$\,\pm\,$}l@{\extracolsep{\fill}}r@{\extracolsep{0pt}}@{$\,\pm\,$}l@{\extracolsep{\fill}}cl}
\hline \hline
R.A.(J2000.0) & Dec.(J2000.0) & \multicolumn{4}{c}{CTIO EC Photometry$^a$} & \multicolumn{2}{c}{Prior cz} & \multicolumn{2}{c}{New cz} & S/N & Notes\\
\cline{3-6}
(h :m :s ) & ($^\circ$ :$^\prime$ :$^{\prime\prime}$ ) & $g$ & $r$ & $i$ & $z$ & \multicolumn{2}{c}{($\mbox{km} \; \mbox{s}^{-1}$)} & \multicolumn{2}{c}{($\mbox{km} \; \mbox{s}^{-1}$)} & & \\[3pt]
\hline \hline\\
\multicolumn{6}{l}{\textbf{NGC 1399 Field}}\\
\multicolumn{6}{l}{\textit{Known Cluster Members}}\\
03:36:26.72 & -35:22:01.6 & 20.20 & 19.42 & 19.11 & 18.79 & 1315&8$^d$   & 1350&27 & 9.5  & UCD-10\\ 
03:36:37.30 & -35:23:09.2 & 20.87 & 20.34 & 20.12 & 19.93 & 1401&13$^d$  & 1383&19 & 8.0  & FCC171 dE galaxy\\ 
03:37:05.75 & -35:37:32.2 & 21.10 & 20.42 & 20.12 & 19.91 & 1520&7$^f$   & 1479&73 & 7.8  & 2dF: background galaxy\\ 
03:37:28.30 & -35:21:22.8 & 20.85 & 20.35 & 20.11 & 19.87 & 1356&139$^c$ & 1373&19 & 6.9  & FCOS 2-2165\\ 
03:37:33.91 & -35:22:19.1 & 20.88 & 20.30 & 20.03 & 19.79 & 2009&106$^c$ & 1997&15 & 10.3 & FCOS 2-2161\\ 
03:38:05.08 & -35:24:09.6 & 19.27 & 18.58 & 18.20 & 17.96 & 1212&32$^b$  & 1219&7  & 20.7 & UCD-6\\ 
03:38:06.33 & -35:28:58.8 & 19.66 & 18.95 & 18.61 & 18.33 & 1234&5$^f$   & 1276&15 & 14.3 & UCD-2\\ 
03:38:10.39 & -35:24:06.1 & 19.98 & 19.32 & 19.00 & 18.76 & 1582&22$^e$  & 1637&15 & 13.5 & UCD-27\\ 
03:38:23.78 & -35:13:49.5 & 20.13 & 19.57 & 19.35 & 19.10 & 1553&72$^c$  & 1637&14 & 9.9  & UCD-32\\ 
03:38:41.98 & -35:33:13.4 & 20.23 & 19.46 & 19.20 & 18.79 & 2010&5$^d$   & 2046&11 & 12.7 & UCD-43; bound to NGC 1404\\ 
03:38:54.10 & -35:33:33.6 & 18.59 & 17.72 & 17.35 & 16.98 & 1491&2$^d$   & 1517&6  & 35.5 & UCD-3\\ 
03:39:35.95 & -35:28:24.5 & 19.64 & 18.88 & 18.58 & 18.22 & 1886&3$^d$   & 1928&11 & 17.0 & UCD-4\\ 
03:39:43.56 & -35:26:59.5 & 20.43 & 19.90 & 19.67 & 19.46 & 1274&7$^f$   & 1250&22 & 7.6  & UCD-54\\ 
03:39:52.58 & -35:04:24.1 & 19.86 & 19.25 & 18.97 & 18.73 & 1245&2$^g$   & 1236&21 & 10.1 & UCD-5\\ 
03:41:29.58 & -35:19:48.8 & --    & --    & --    & --    & 1473&6$^f$   & 1455&89 & 7.3  & gc129.2; outside CTIO fields\\ 
\\
\multicolumn{6}{l}{\textit{New CSSs}}\\
03:38:34.04 & -35:25:22.0 & 22.11 & 21.65 & 21.37 & 21.24 & \multicolumn{2}{c}{--} & 1478&32 & 13.4 & \parbox{5cm}{point source at R.A. 03:38:33.80 Dec. -35:25:21.5 in CTIO e1n1 image from Karick} \\[6pt] 
03:43:08.16 & -35:23:43.4 & -- & -- & -- & -- & \multicolumn{2}{c}{--} & 1271&24 & 5.8  & \parbox{5cm}{outside CTIO fields; $10^\prime$E of NGC 1427}\\ 
\\
\multicolumn{6}{l}{\textbf{NGC 1316 Intracluster/Galaxy Merger Field}}\\
\multicolumn{6}{l}{No cluster members located.} \\
\\
\hline
\end{tabular*}
}
\begin{flushleft}
\item a. Error-corrected photometry where available from CTIO \citep{Karick..2007}.
\item b. Redshifts from 2dF surveys \citep{Drinkwater..2000a, Gregg..2007}.\\
\item c. Fornax Compact Object Survey \citep{Mieske..2004I}.\\
\item d. From our observations with VLT \citep{Firth..2007}.\\
\item e. \citet{Dirsch..2004}.\\
\item f. \citet{Bergond..2007}.\\
\item g. \citet{Hilker..2007}.\\
\end{flushleft}
\end{table*}

\begin{figure}
\centering
 \includegraphics[width=8.6cm]{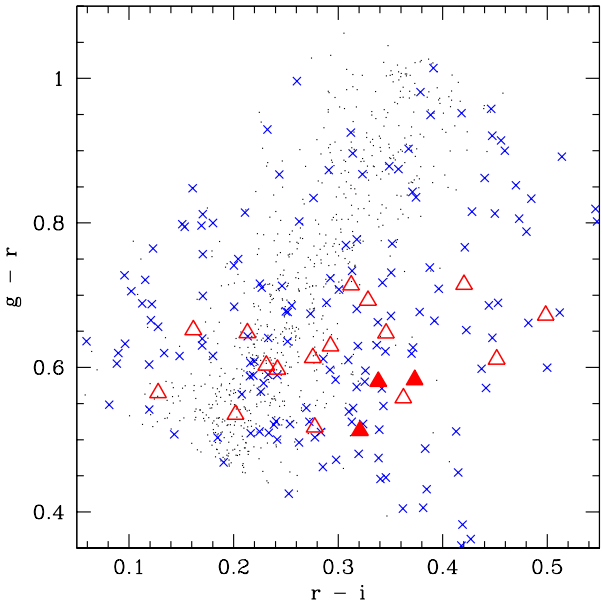}
 \caption{Colour-colour plot of AAOmega Virgo redshifts for Galactic stars (dots), background galaxies (crosses), cluster core CSSs (open triangles) and IGC's (filled triangles). The colour locus of Galactic stars slopes from lower left to upper right, and the rectangular target selection box is apparent. The three Virgo IGC's have similar colours.}
 \label{fig:Virgogri}
 \end{figure}

As already explained, for each cluster we used a different method of reducing the large numbers of point source targets in our search for CSSs. The results of our colour-colour selection method in the Virgo Cluster are evident in Fig.~\ref{fig:Virgogri} which compares the distribution of CSSs with that of foreground stars and background galaxies located by our observations. The Galactic stellar locus evident in both Virgo fields contrasts with the widely distributed background galaxy colours. CSSs detected in our survey are not distributed along the stellar locus. Based on the positions of detected CSSs, we could improve our targeting efficiency in future surveys by restricting targets to $g - r \le 0.8$.

\begin{figure}
\centering
 \includegraphics[width=8.6cm]{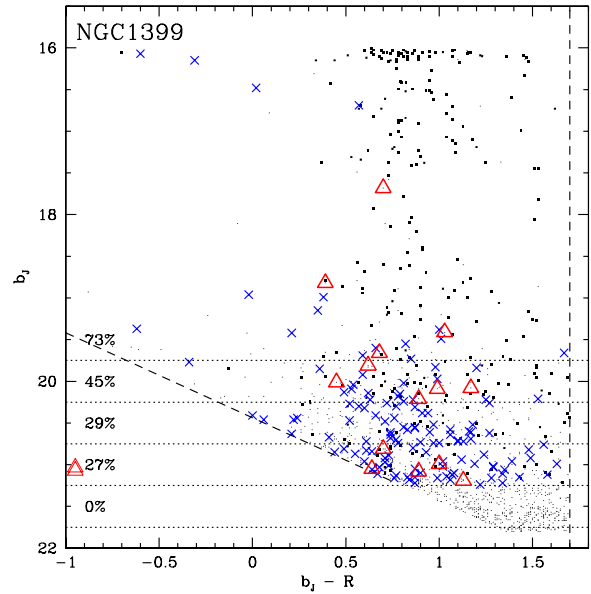}
 \caption{Colour-magnitude plot of NGC 1399 targets (dots) and redshifts obtained for Galactic stars (small squares), background galaxies (crosses) and CSSs (triangles). CSSs with invalid $R$-band photometry are shown adjacent to the magnitude axis. NGC 1399 targets were selected from deep CCD imaging \citep{Karick..2005} with the CTIO telescope; the vertical and inclined dashed lines are the colour selection and $R<20.4$ plate detection limits respectively.}
 \label{fig:Fornax_52}
 \end{figure}

In the Fornax Cluster, Fig.~\ref{fig:Fornax_52} shows the results of the colour-magnitude selection method we adopted. In the previously observed NGC 1399 field we extended our existing 2dF redshift data to the upper end of the GC distribution at $b_J \le 21.2$ ($M_{b_J} \le -10.2$), whereas for the NGC 1316 field with limited prior point-source redshift data we concentrated on the brighter targets to obtain $79\%$ completeness for $b_J \le 19.7$ ($M_{b_J} \le -11.7$ -- this faint limit can only locate massive, bright CSSs).
 
\begin{figure*}
\centering
 \includegraphics[width=8.5cm]{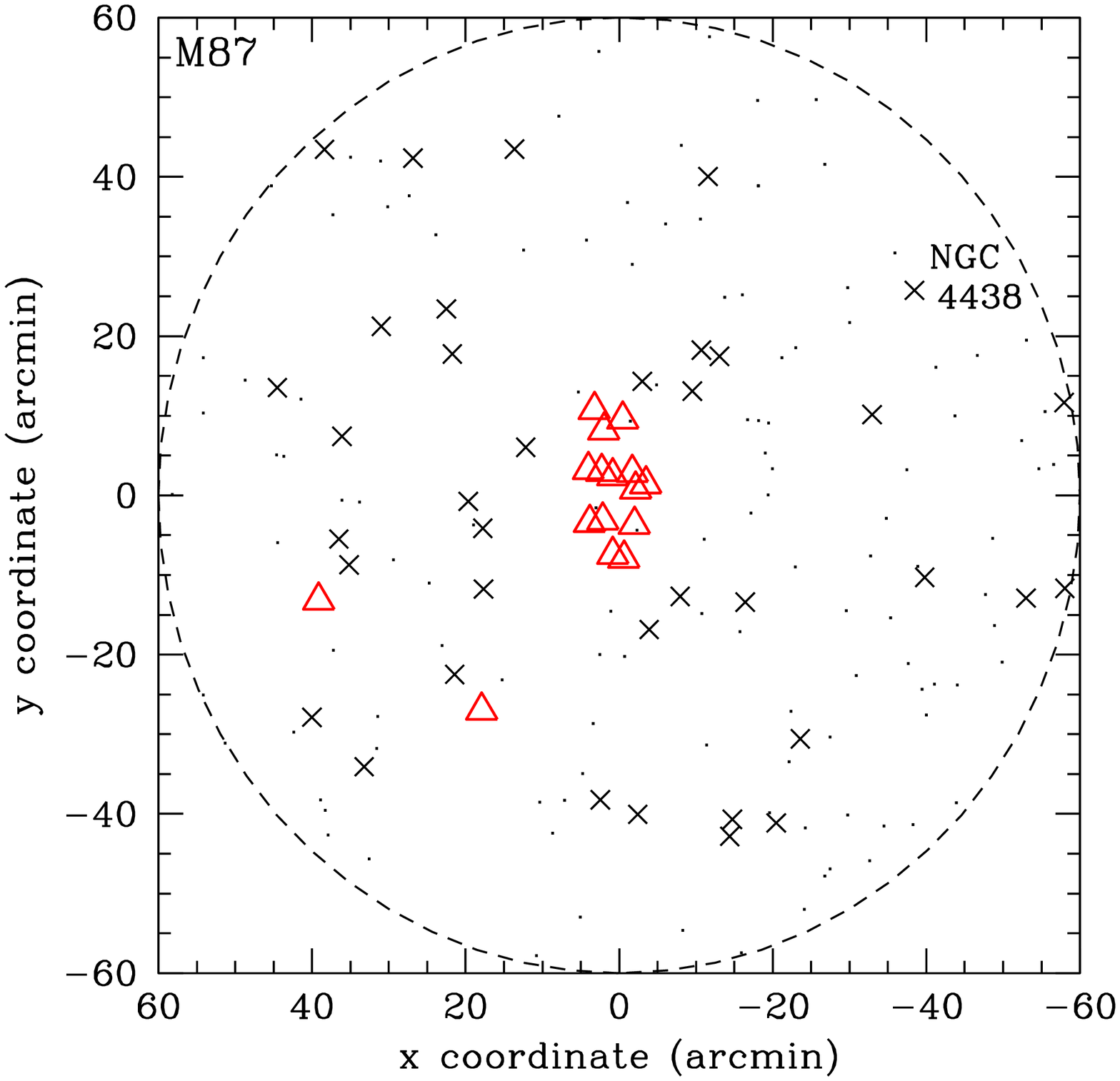}
 \includegraphics[width=8.5cm]{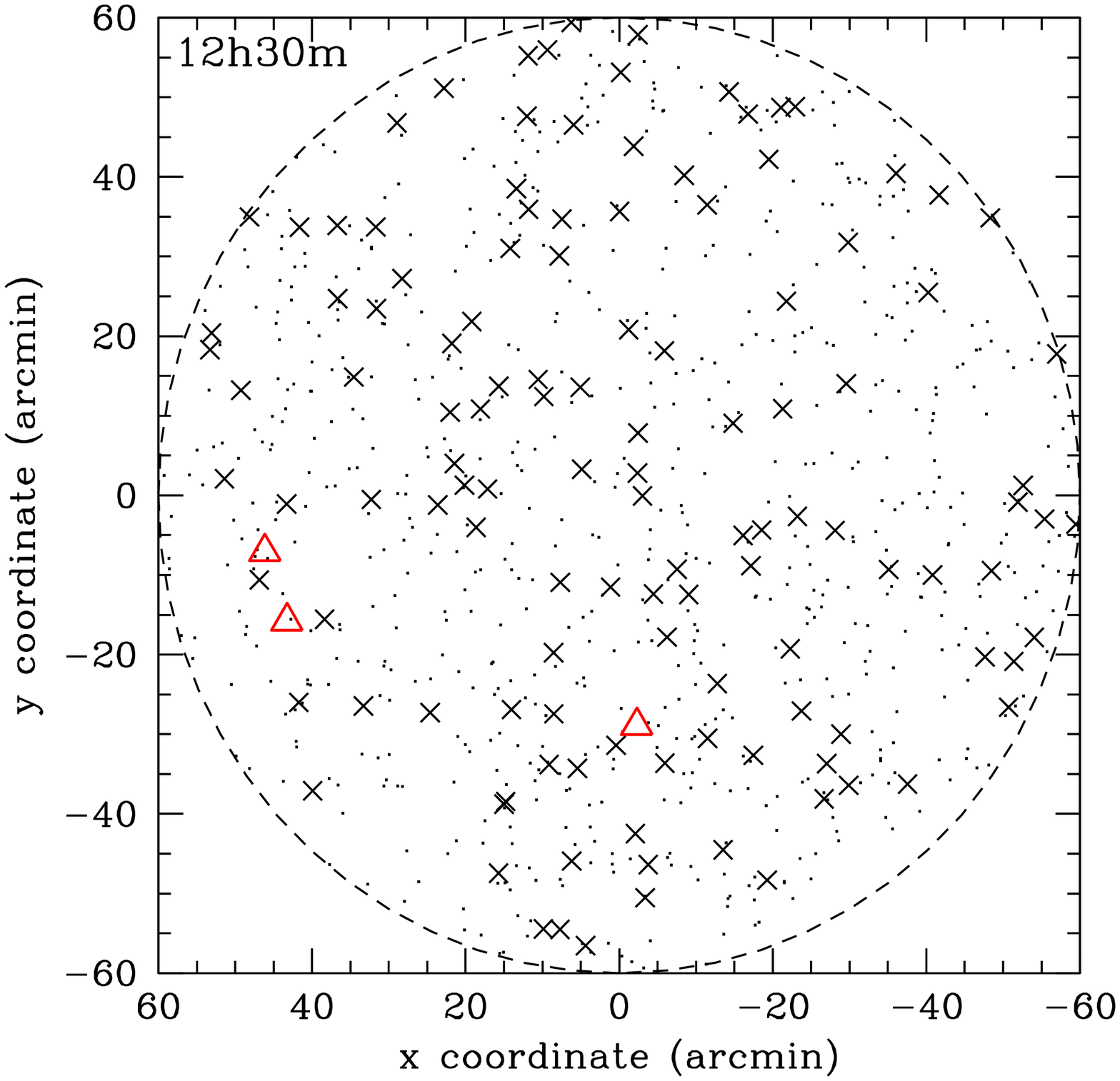}
 \vspace{0.5cm}
 \includegraphics[width=8.5cm]{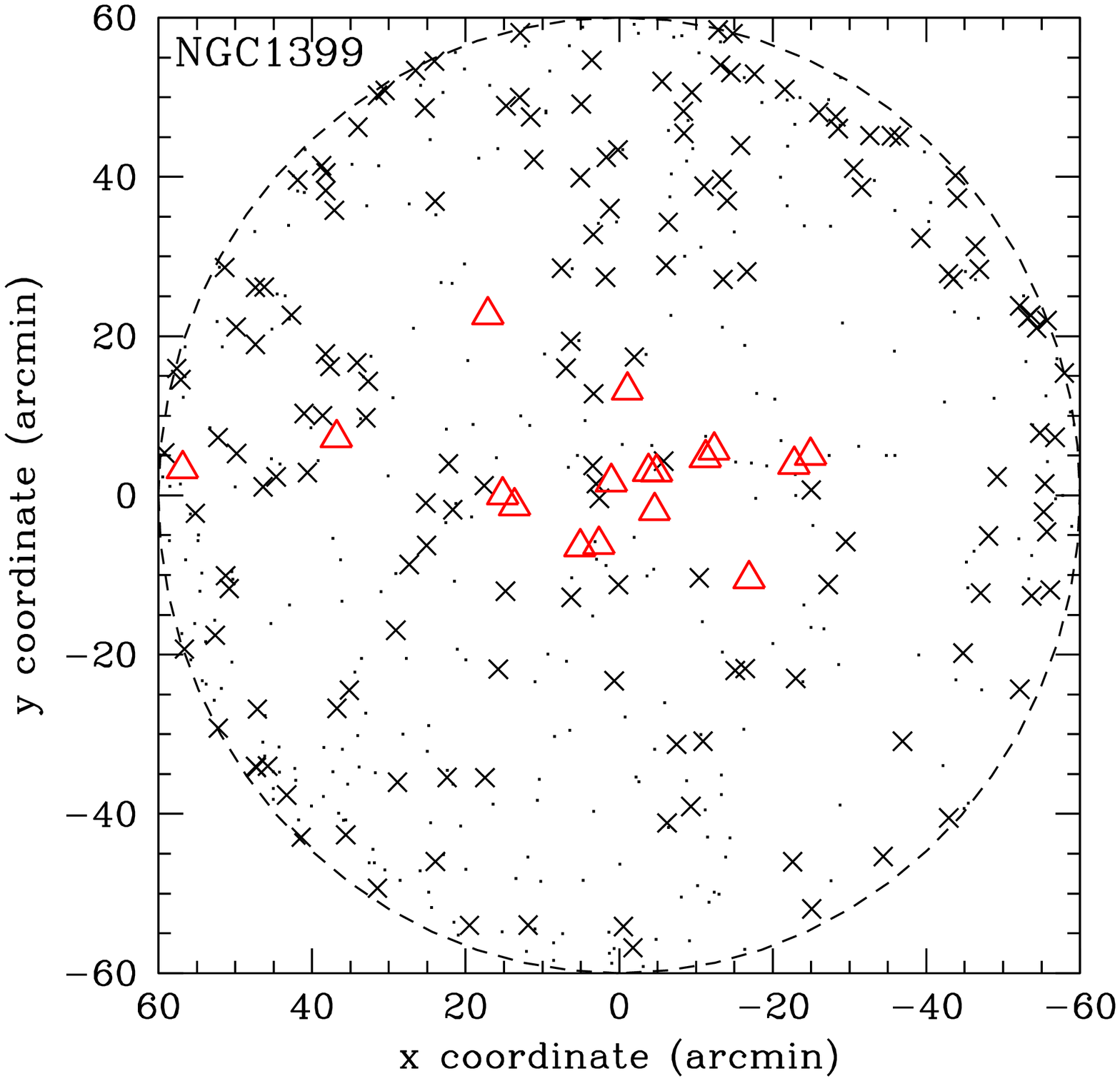}
 \includegraphics[width=8.5cm]{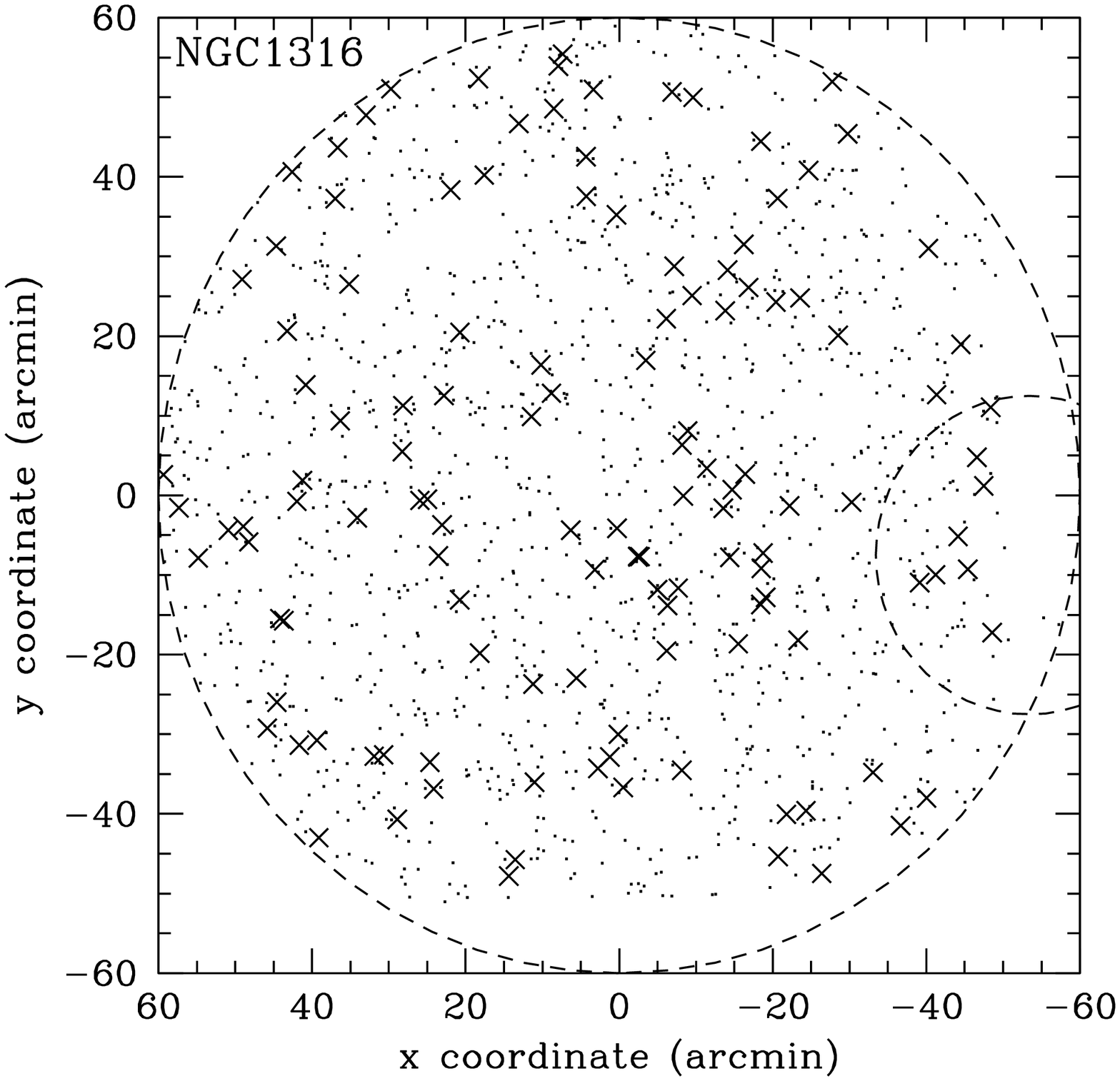}
 \caption{Redshift results for the Virgo (upper) and Fornax (lower) cluster fields, distinguishing between background galaxies (crosses), cluster objects (triangles) and Galactic stars (dots: cz$< 500 \, \mbox{km} \; \mbox{s}^{-1}$). CSSs are concentrated near M87 and NGC 1399, but are dispersed or absent in the intracluster fields. The interacting galaxy NGC 4438 at the edge of the M87 field does not have an associated overdensity of confirmed foreground or cluster point sources located in our survey. In the Fornax intracluster field, the galaxy merger environment is indicated by a 20 arcmin ($130 \, \mbox{kpc}$) dashed arc centred on NGC 1316.}
 \label{fig:aaovirgofornax_11}
 \end{figure*}

In line with previous 2dF results \citep{Drinkwater..2000a, Jones..2006, Gregg..2007}, we find that CSSs are centrally concentrated in the cluster core M87 and NGC 1399 fields, as illustrated in Fig.~\ref{fig:aaovirgofornax_11}. The newly located IGCs appear more widely dispersed across the Virgo intracluster field, but this may be due to our limited sampling of the IGC population. There is no clear evidence of background galaxy clustering in our observation results.
 
\begin{figure*}
\centering
   \includegraphics[width=8.0cm]{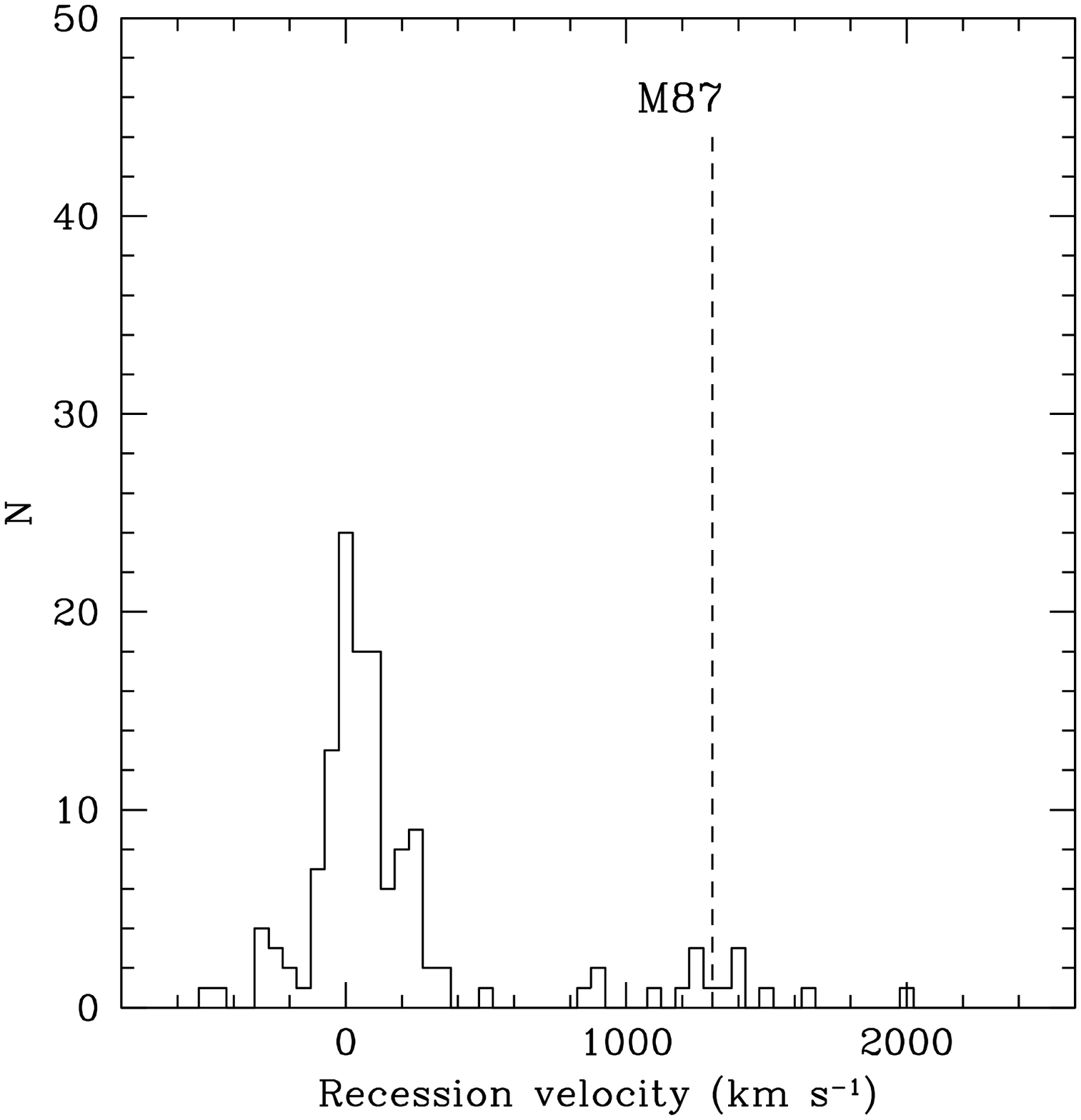}
   \includegraphics[width=8.0cm]{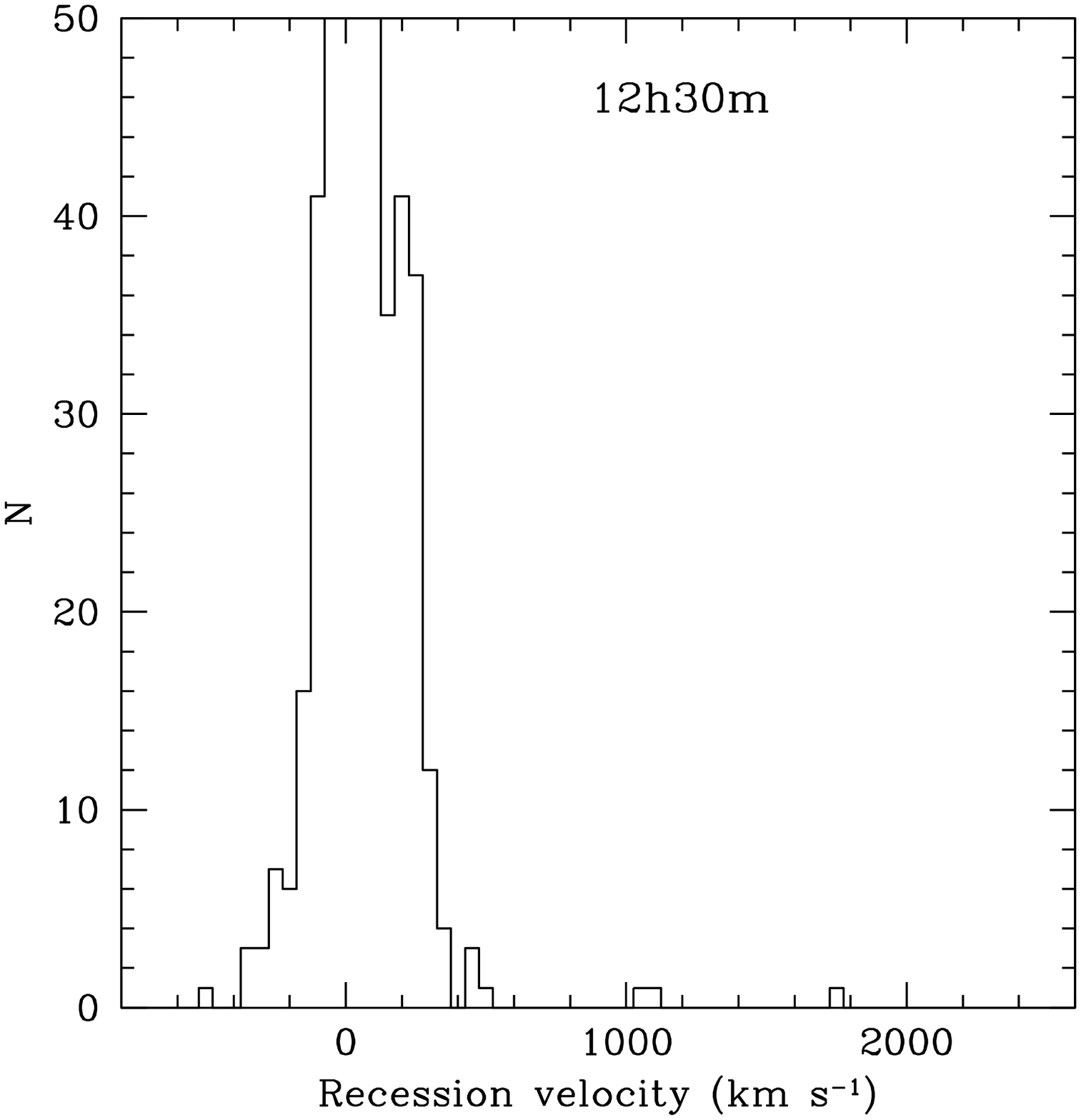}
   \includegraphics[width=8.0cm]{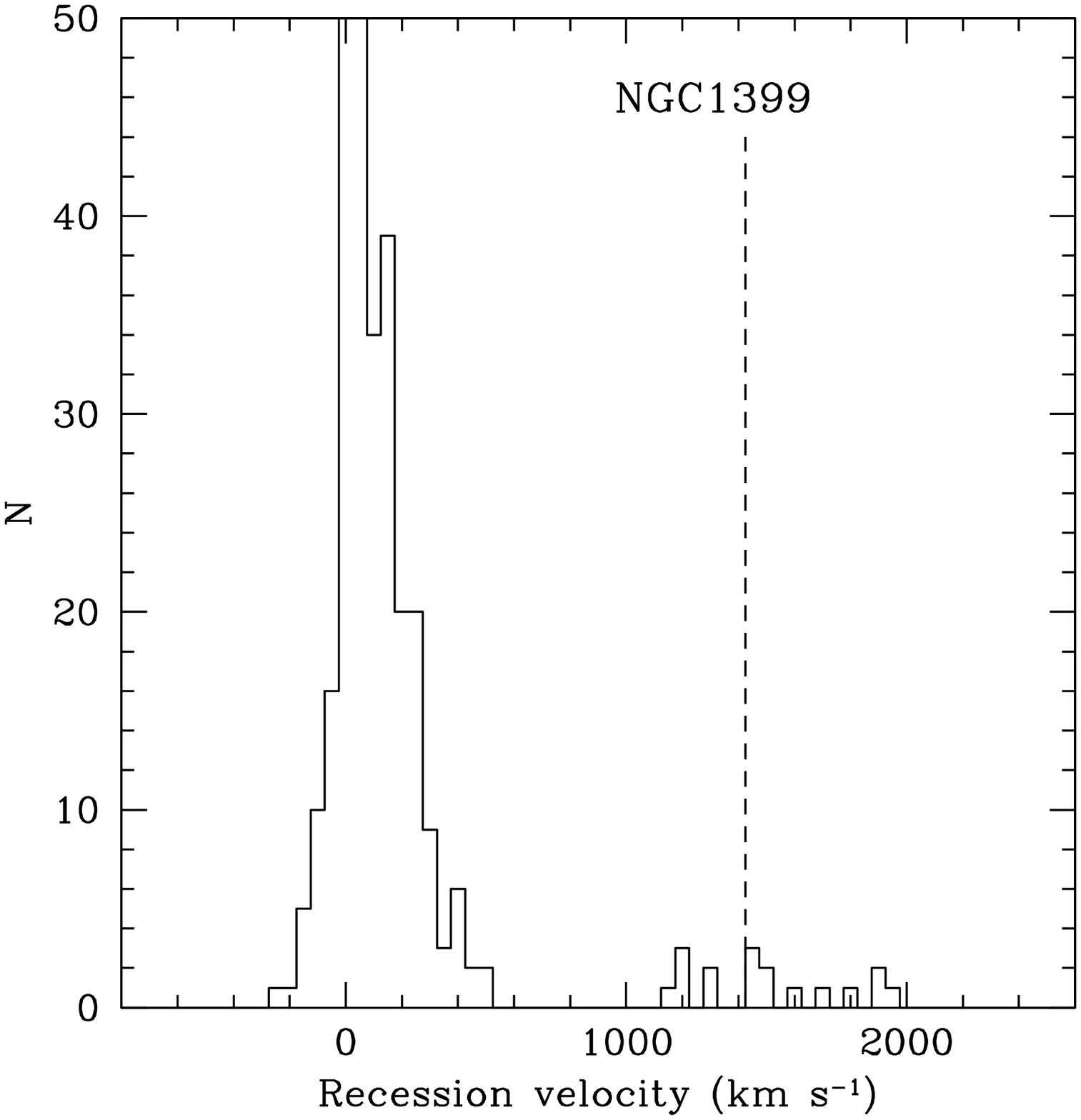}
   \includegraphics[width=8.0cm]{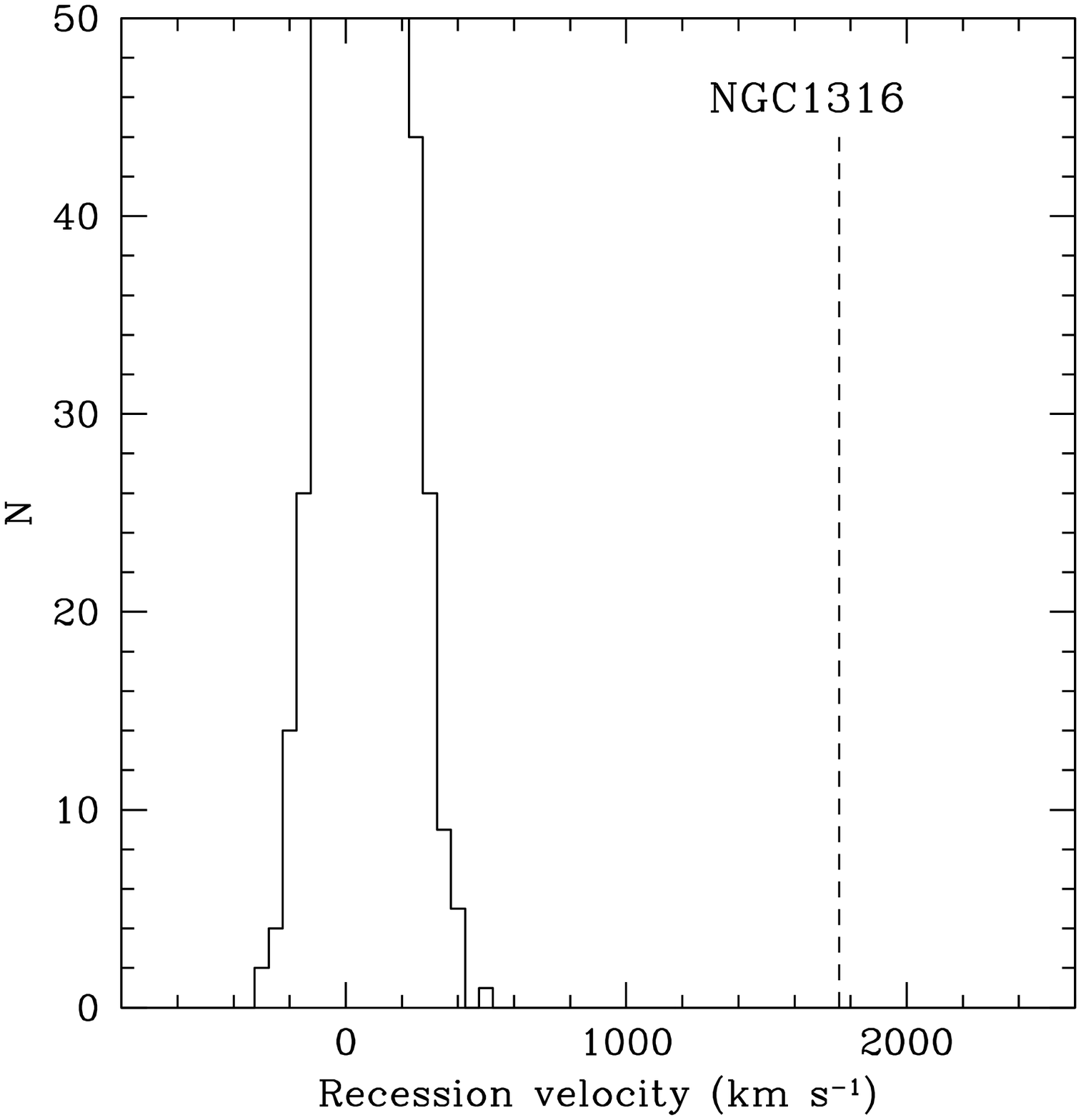}
 \caption{Near-field recession velocity histograms for each Virgo and Fornax cluster field. The near-field is dominated by Galactic stars. While Fornax is isolated in redshift space, Virgo and the Galaxy are part of the same supercluster, so we see a less obvious transition between stars and cluster members. The small bumps in the Virgo distribution at approximately -200 to -400$\; \mbox{km} \, \mbox{s}^{-1}$ and 200 to 500$\; \mbox{km} \, \mbox{s}^{-1}$ (particularly noticeable in the M87 field but note the effect of small number statistics) may be partly due to Virgo cluster members with low or negative recession velocities.}
 \label{fig:aaovirgofornax_12}
 \end{figure*}

We attempt to separate cluster objects from foreground stars in Fig.~\ref{fig:aaovirgofornax_12}, which plots the heliocentric recession velocity distributions of our results in near-field ($<\!2600 \; \mbox{km} \, \mbox{s}^{-1}$) redshift space. Galactic stars dominate our results at low redshift, particularly in the previously unobserved intracluster fields. At higher redshift we preferentially detect background emission-line galaxies and QSOs, which retain prominent emission features in low S/N spectra.

\section{Analysis and Discussion}

\subsection{Cumulative CSS Datasets}
Our aim is to compare the spatial distribution, kinematics and colours of spectroscopically confirmed Virgo and Fornax CSSs in different cluster environments to the predictions of CSS formation theories. To provide the largest possible spectroscopically confirmed dataset for this comparison, we have combined bright CSSs detected in our AAOmega observations with previously catalogued GCs, dEs and bright CSSs from the sources listed below. We will use the combined M87 and NGC 1399 datasets in subsection \ref{clustercore}, and the NGC 1316 dataset in subsection \ref{galaxymerger}.

\begin{itemize}
\item \textbf{M87:} (a) Using 2dF \citet{Jones..2006} located 9 UCDs in the core region of the Virgo Cluster through a survey of colour-selected point sources ($b_J<21.5$). More precise recession velocities, together with other properties, were subsequently published for six of these UCDs by \citet{Evstigneeva..2007}. (b) The key source of published redshift data for the M87 GC system is the survey by \citet{Hanes..2001} confirming 286 GCs with recession velocities between 500 and 2500 $\mbox{km} \, \mbox{s}^{-1}$. In addition, \citet{Schuberth..2006} has published redshift data for 174 GCs located between 0.9 and $15.5 \; \mbox{arcmin}$ from M87. 
\item \textbf{NGC 1399:} (a) 60 UCDs were located with 2dF in the all-object Fornax Cluster Spectroscopic Survey \citep{Drinkwater..2000b, Phillipps..2001} to $b_J<19.7$, and in a follow-up survey of fainter ($b_J<21.5$) blue point sources in the core region \citep{Drinkwater..2005, Gregg..2007}. More precise recession velocities, were subsequently published for four of these UCDs using the UVES-VLT spectrograph \citep{Hilker..2007}. (b) The dataset comprising 468 redshifts published by \citet{Dirsch..2004} is our key source for the inner (2 to 9$\; \mbox{arcmin}$) GC system of NGC 1399. Approximately 160 additional redshifts of the outer GC system were described by \citet{Schuberth..2004} but the dataset and positions have not yet been published. \citet{Bergond..2007} has published redshifts for 149 GCs, including 61 that are potentially IGCs. However, by using photometric data \citet{Schuberth..2008} contends that all but one of these IGCs are either bound to a local host galaxy or part of the extended GC system of NGC 1399.
\item\textbf{NGC 1316:} \citet{Goudfrooij..2001a} has published a catalog of 24 CSSs, including 10 with $M_V<-10.8$, found by multi-slit spectroscopy within an $8 \times 8 \, \mbox{arcmin}^{-2}$ field centred on NGC 1316.
\end{itemize}

In the following analysis we avoid sampling bias when comparing between Virgo and Fornax or extrapolating results to the overall CSS population, by extracting subsets common to the different sampled parameter spaces (colours, magnitudes and areal coverage), as described below.

\subsection{The Cluster Core Environment}
\label{clustercore}
In this subsection we analyse the central CSS populations and in particular compare the Fornax and Virgo populations.

\subsubsection{Bright CSS Populations in Virgo and Fornax}
\begin{figure*}
\centering
 \includegraphics[width=8.5cm]{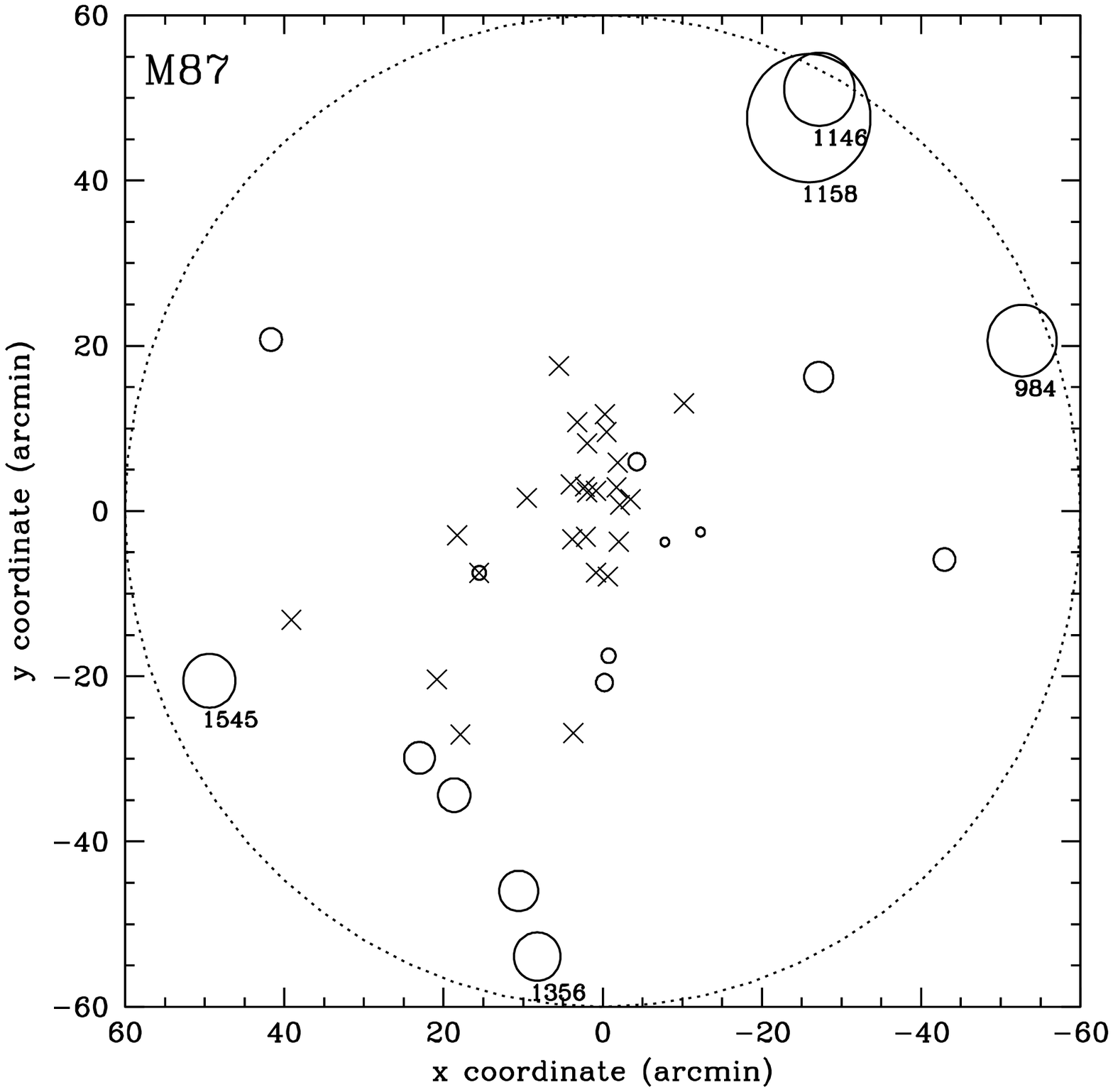}
 \includegraphics[width=8.5cm]{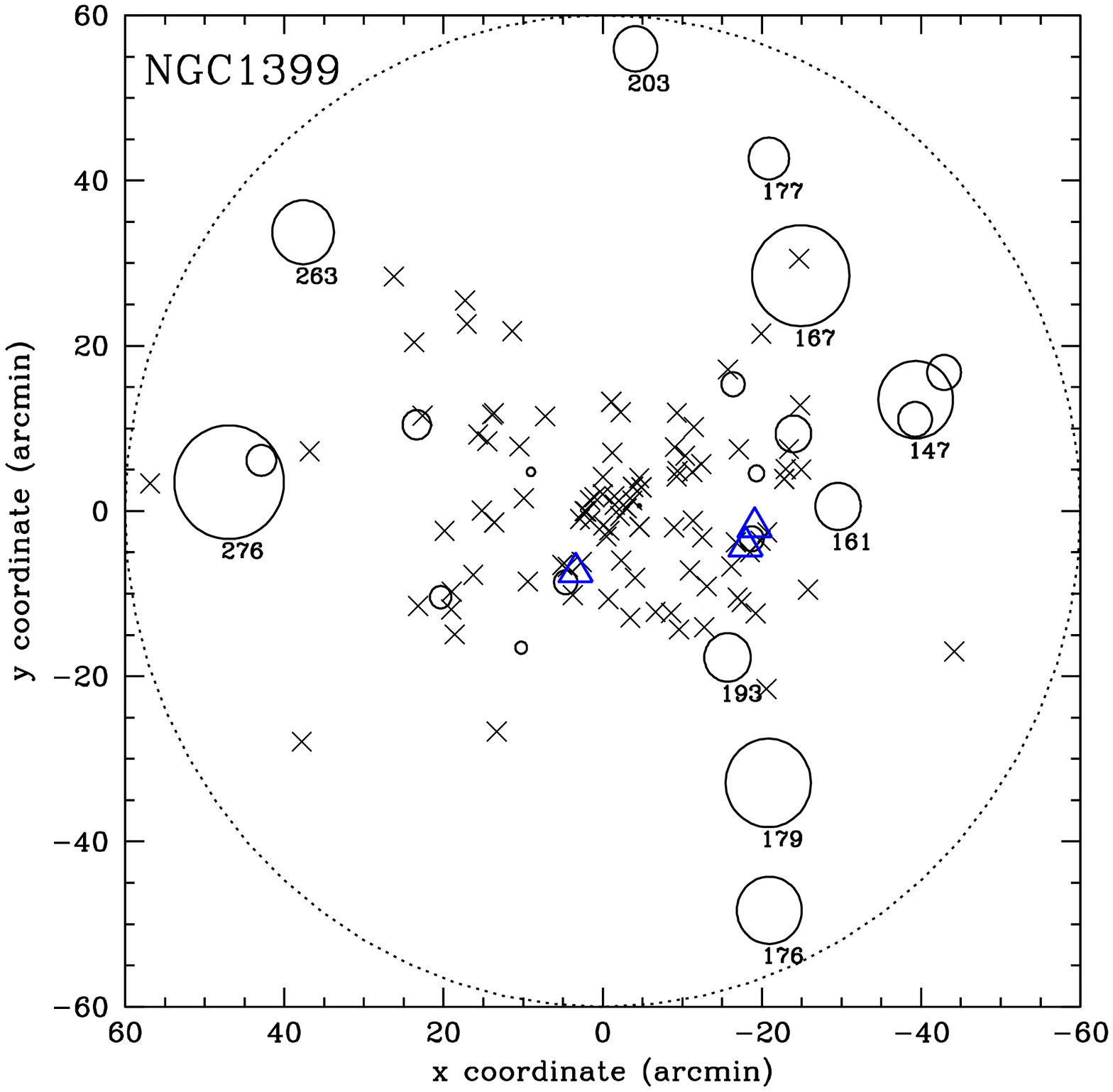}
\caption{Distribution of bound CSSs (triangles) and unbound CSSs (crosses) within the Virgo and Fornax cluster core environments. Tidal radii (circles) of non-dwarf galaxies are computed with respect to M87 and NGC 1399 respectively. Identifying numbers for the galaxies with larger tidal radii are from the Virgo and Fornax Cluster Catalogues. In intracluster environments, tidal radii are meaningless as an indicator of bound CSSs.}
\label{fig:aao_1010}
\end{figure*}

To better understand their origins, we firstly identify luminous CSSs that are gravitationally bound to non-dwarf satellite galaxies of M87 or NGC 1399 (see Fig.~\ref{fig:aao_1010}). In the cluster core regions dominated by giant elliptical galaxies, CSSs are considered to be bound objects if they are within the tidal radius of a non-dwarf galaxy and have a relative recession velocity less than the escape velocity \citep[see][for a description of this method]{Firth..2007}. Since only 4 out of 105 luminous CSSs in the NGC 1399 cluster core region (and none in the M87 region) are bound to satellite galaxies, our results are not significantly affected by their inclusion.
  
Table \ref{table:csscore} combines previous 2dF surveys \citep{Drinkwater..2000a, Gregg..2007, Jones..2006} with our AAOmega surveys, and compares in absolute magnitude bins the number of luminous CSSs located in the core regions surrounding M87 and NGC 1399, separating those CSS that are brighter ($M_{b_J} \le -10.5$) than the typical magnitude range of GCs. After eliminating target duplications, we calculate completeness with respect to the APM catalogue of point sources within the respective cluster core fields and colour-magnitude limits ($15.0 < b_J < 21.8$, $b_J - R \le 1.6$). The Palomar survey plates used for the Virgo APM catalogue have a shallower faint limit than the UKST survey plates used for the Fornax APM catalogue, so APM point sources in our M87 catalogue tail off for $M_{b_J}>-10.5$, but we restrict our CSS population estimates to the luminous CSSs. Both cluster core fields in Table \ref{table:csscore} have the same spatial extent ($\sim\!630 \; \mbox{kpc}$) -- the $1.8^\circ$-diameter field for the more distant NGC 1399 is spatially equivalent to the $2^\circ$-diameter M87 field.

Within the magnitude range $M_{b_J}<-10.5$ we estimate, by adjusting for completeness and using Poisson statistics for uncertainty estimation, that M87 has $25\pm8$ luminous CSSs, or half the number ($47\pm12$) estimated for NGC 1399. This finding contrasts sharply with the relative size of the innermost GC populations, usually expressed as the luminosity-scaled specific frequency $S_N$\footnote{$S_N = N_{GC} 10^{0.4(M_V+15)}$, where $N_{GC}$ is the estimated number of GCs and $M_V$ is the galaxy absolute magnitude.} \citep[see review by][]{Elmegreen..1999}. Based on observations within $7 \; \mbox{arcmin}$ of M87 and NGC 1399, \citet{Forte..2002} calculated $S_N$ of $6\pm1$ and $3.7\pm0.8$, galaxy luminosities of $V=8.54\pm0.01$ and $V=9.02\pm0.06$, and total GC populations of $4700\pm400$ and $2300\pm300$ respectively. If luminous CSSs are simply an extension of the central giant elliptical galaxy GC population, we would expect the Virgo cluster core to have a luminous CSS population $\sim4$ times greater than we have estimated from our redshift surveys. Even though the $S_N$ estimates are not redshift confirmed we consider they are likely to be accurate within a factor of $\sim\!2$, so our contrary estimates for the luminous CSS populations may be due either to a real difference with the GC populations or to uncertainties in population estimates caused by lack of observing completeness in Virgo (see Table \ref{table:csscore}) -- further observations would clarify this.

\begin{table*}
\caption{Cluster Core Luminous CSS Results}
\label{table:csscore}
\centering
\scriptsize{
\begin{tabular*}{1.00\textwidth}
     {@{\extracolsep{\fill}}ccccccc}
\hline \hline
$M_{b_J}$ Range & APM & \multicolumn{2}{c}{Redshifts} & Completeness & \multicolumn{2}{c}{CSSs Found}\\
\cline{3-4} \cline{6-7}
(mag) & Targets & 2dF$^a$ & AAOmega$^b$ & (per cent) & 2dF & AAOmega\\[3pt]
\hline \hline\\
\multicolumn{6}{l}{M87 -- $2^\circ$ Field}\\
 -14.0 to -14.5 & 344 & 66  &    & 19.2 & 0 & -- \\
 -13.5 to -14.0 & 373 & 73  & 21 & 25.2 & 0 & 0 \\
 -13.0 to -13.5 & 399 & 132 & 8  & 35.1 & 1 & 0 \\
 -12.5 to -13.0 & 503 & 238 & 5  & 48.3 & 0 & 0 \\
 -12.0 to -12.5 & 537 & 253 & 3  & 47.7 & 0 & 0 \\
 -11.5 to -12.0 & 505 & 301 & 6  & 60.8 & 1 & 0 \\
 -11.0 to -11.5 & 430 & 240 & 15 & 59.3 & 5 & 2 \\
 -10.5 to -11.0 & 345 & 229 & 31 & 75.4 & 3 & 1 \\
 \\
 -10.0 to -10.5 & 18  &     & 54 & --   & -- & 5 \\ 
 -9.5  to -10.0 & 0   &     & 30 & --   & -- & 8 \\
\\
\multicolumn{6}{l}{NGC 1399 -- $1.8^\circ$ Field}\\
 -15.0 to -15.5 & 183  & 82  & 34  & 63.4  & 0  & 0 \\
 -14.5 to -15.0 & 225  & 214 & 22  & 100.0 & 0  & 0 \\
 -14.0 to -14.5 & 281  & 255 & 14  & 95.7  & 0  & 0 \\
 -13.5 to -14.0 & 328  & 319 & 17  & 100.0 & 0  & 0 \\
 -13.0 to -13.5 & 347  & 341 & 5   & 99.7  & 0  & 0 \\
 -12.5 to -13.0 & 466  & 458 & 7   & 99.8  & 0  & 0 \\
 -12.0 to -12.5 & 588  & 575 & 12  & 99.8  & 0  & 0 \\
 -11.5 to -12.0 & 758  & 673 & 16  & 90.9  & 1  & 0 \\
 -11.0 to -11.5 & 1040 & 705 & 23  & 70.0  & 4  & 0 \\
 -10.5 to -11.0 & 1474 & 642 & 33  & 45.8  & 17 & 0 \\
 \\
 -10.0 to -10.5 & 1819 & 376 & 80  & 25.1  & 15 & 5 \\ 
 -9.5  to -10.0 & 2506 & 71  & 33  & 4.2   & 2  & 1 \\ 
\hline
\end{tabular*}
}
\begin{list}{\hspace{0.5 cm}}{}
\item a. Redshifts from 2dF surveys \citep{Drinkwater..2000a, Gregg..2007}.\\
\item b. Some AAOmega redshifts duplicate 2dF redshifts, but this has been corrected in our completeness percentages.\\
\item c. We separate the absolute magnitude range $M_{b_J} \le -10.5$ covering the luminous CSS and for which we have a reasonable level of redshift completeness in both galaxy clusters.\\
\end{list}
\end{table*}

\subsubsection{Recession Velocity Dispersion}
\begin{figure*}
\centering
\includegraphics[width=15cm]{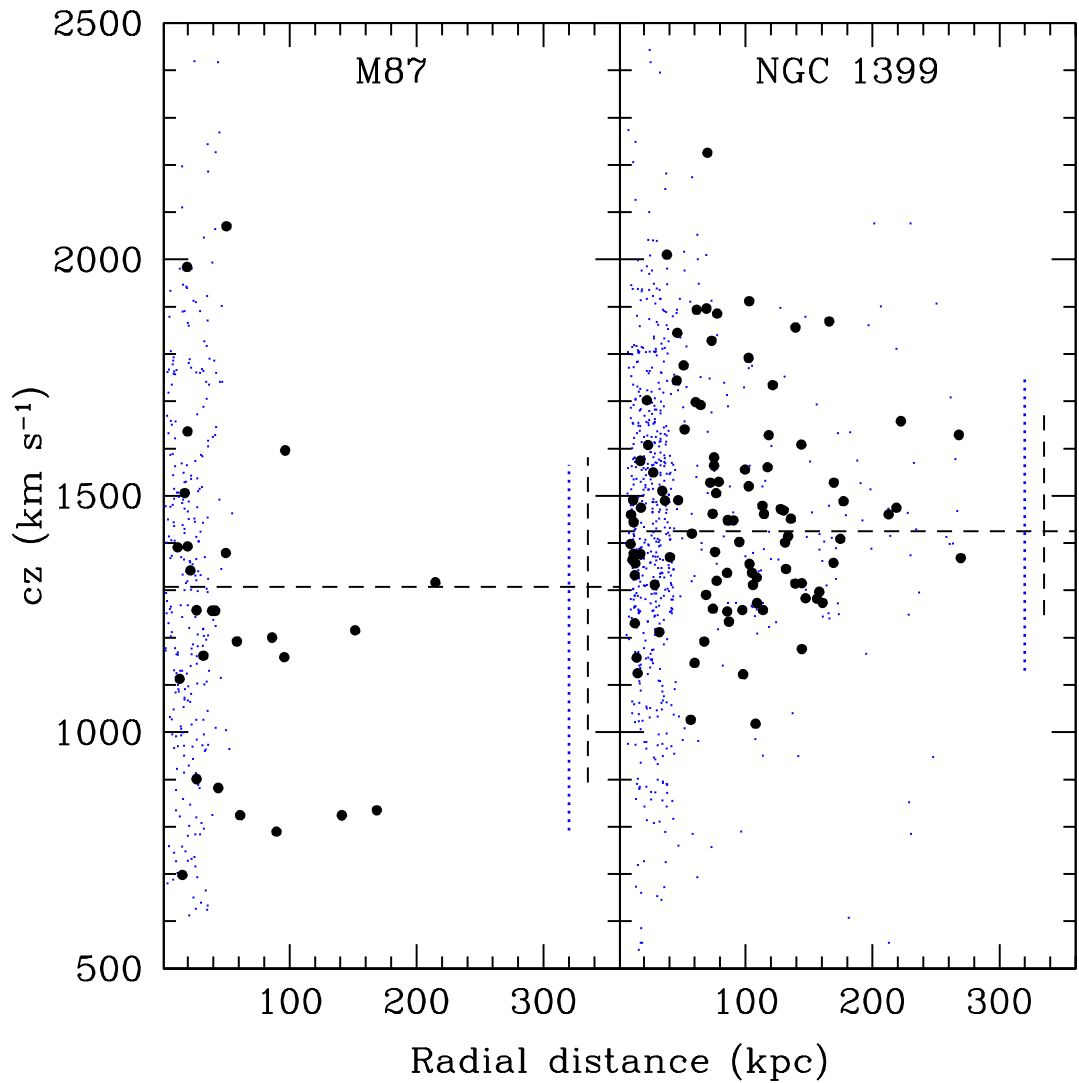}
\caption{CSS recession velocity distribution as a function of radial distance from the cD galaxy. GCs (points) are compared with bright/spatially dispersed CSSs (filled circles). The horizontal dashed lines mark cD galaxy recession velocities. Vertical lines show $\pm1\sigma$ velocity dispersions of GCs (dotted line) and luminous CSSs (dashed line). \textsc{Left:} The Virgo Cluster core (M87) luminous CSS velocity dispersion is $344\pm15 \; \mbox{km} \, \mbox{sec}^{-1}$. \textsc{Right:} The Fornax Cluster core (NGC 1399) luminous CSS velocity dispersion is $228\pm34 \; \mbox{km} \, \mbox{sec}^{-1}$.}
\label{fig:combinedcz_1030}
\end{figure*}

Recession velocity dispersion is a useful method to compare CSS sub-populations that may have differing kinematical histories -- for example, how do the velocities of luminous CSSs (at least two or three magnitudes brighter than the peak of the GC luminosity function) compare with their fainter counterparts? Table \ref{table:velocities} and Fig.~\ref{fig:combinedcz_1030} compare the heliocentric recession velocity distributions of GCs and luminous CSSs in the Fornax and Virgo cluster core regions.

Firstly, the larger velocity dispersions of both GCs and luminous CSSs in Virgo shows that both sub-populations are dynamically more energetic in Virgo than in Fornax, as confirmed by the statistical test results shown in Table \ref{table:velocities}. These results are expected since M87 is more massive than NGC 1399, and embedded in a more complex cluster undergoing relaxation. Secondly, in each cluster the velocity dispersion of luminous CSSs is less than that of the central galaxy's GC system; however, this difference is more significant in Fornax ($<0.01$ $F$-test probability that this is due to sampling variability) than in Virgo (0.49 $F$-test probability that this is due to sampling variability).

These results provide clear evidence in Fornax that luminous CSSs form a dynamically distinct population to the central galaxy GCs. A possible explanation is that in Fornax the majority of luminous CSSs formed from tidally stripped dwarf galaxies in preferentially less energetic orbits \citep[as discussed by][]{Bekki..2001, Bekki..2003I}.

\begin{table*}
\caption{Cluster Core Recession Velocity Data}
\label{table:velocities}
\centering
\scriptsize{
\begin{tabular*}{1.00\textwidth}
     {@{\extracolsep{\fill}}lcccccccccc}
\hline \hline
\vspace{2pt}
 & \multicolumn{6}{c}{Mean Velocity} & & \multicolumn{3}{c}{Velocity Dispersion}\\
\cline{2-7} \cline{9-11}
 & \multicolumn{2}{c}{GCs} & & \multicolumn{2}{c}{Luminous CSSs} & $t$-test$^c$ & & GCs & Luminous CSSs & $F$-test$^d$\\
 & (N) & ($\mbox{km} \, \mbox{s}^{-1}$) & & (N) & ($\mbox{km} \, \mbox{s}^{-1}$) & Probability &  & ($\mbox{km} \, \mbox{s}^{-1}$) & ($\mbox{km} \, \mbox{s}^{-1}$) & Probability \\[3pt]
\hline \hline\\
Fornax (NGC 1399) & 758 & $1443\pm37^a$ & & 109 & $1475\pm35$ & 0.19 & & $313\pm40^a$ & $228\pm34$ & $<0.01$ \\
\\
Virgo (M87) & 455 & $1179\pm74^b$ & & 26 & $1238\pm38$ & 0.41 & & $387\pm39^b$ & $344\pm15$ & 0.49 \\
\\
Comparison Tests & & $<0.01$ $^c$ & & & $<0.01$ $^c$ &  &  & $<0.01$ $^d$ & $<0.01$ $^d$ &  \\[3pt]
\hline
\end{tabular*}
}
\begin{flushleft}
\item a. GC data extracted from \citet*{Mieske..2002}; \citet{Dirsch..2004}; \citet{Mieske..2004I, Bergond..2007}.\\
\item b. GC data extracted from \citet{Hanes..2001, Schuberth..2006}.\\
\item c. The $t$-test probability tests the hypothesis that two sample means are drawn from the same population.\\
\item d. The $F$-test probability tests the hypothesis that two sample variances are drawn from the same population.\\
\end{flushleft}
\end{table*}

\subsubsection{CSS Photometric Properties}
To compare CSS colours and magnitudes in the core regions of Virgo and Fornax, we need to convert our results to the same extinction-corrected (de-reddened) photometry. We therefore obtained $gri$ de-reddened Virgo photometry from SDSS \citep{Adelman..2006}\footnote{The final SDSS $ugriz$ photometry obtained with the 2.5-m survey telescope at Apache Point Observatory should not be confused with the $u^\prime g^\prime r^\prime i^\prime z^\prime$ system \citep{Fukugita..1996} which refers to filter/grating-detector combinations mounted on other telescopes (such as the SDSS $20^{\prime\prime}$ photometric monitoring telescope).} and cross-matched our Fornax CSS results to de-reddened CTIO photometry \citep{Karick..2007} covering most of the NGC 1399 one square degree field, having checked that these two photometry sources are well calibrated. We were unable to obtain CTIO photometry for 10 Fornax CSSs and have excluded them from the following analysis.

\begin{figure}
\centering
\includegraphics[width=8.5cm]{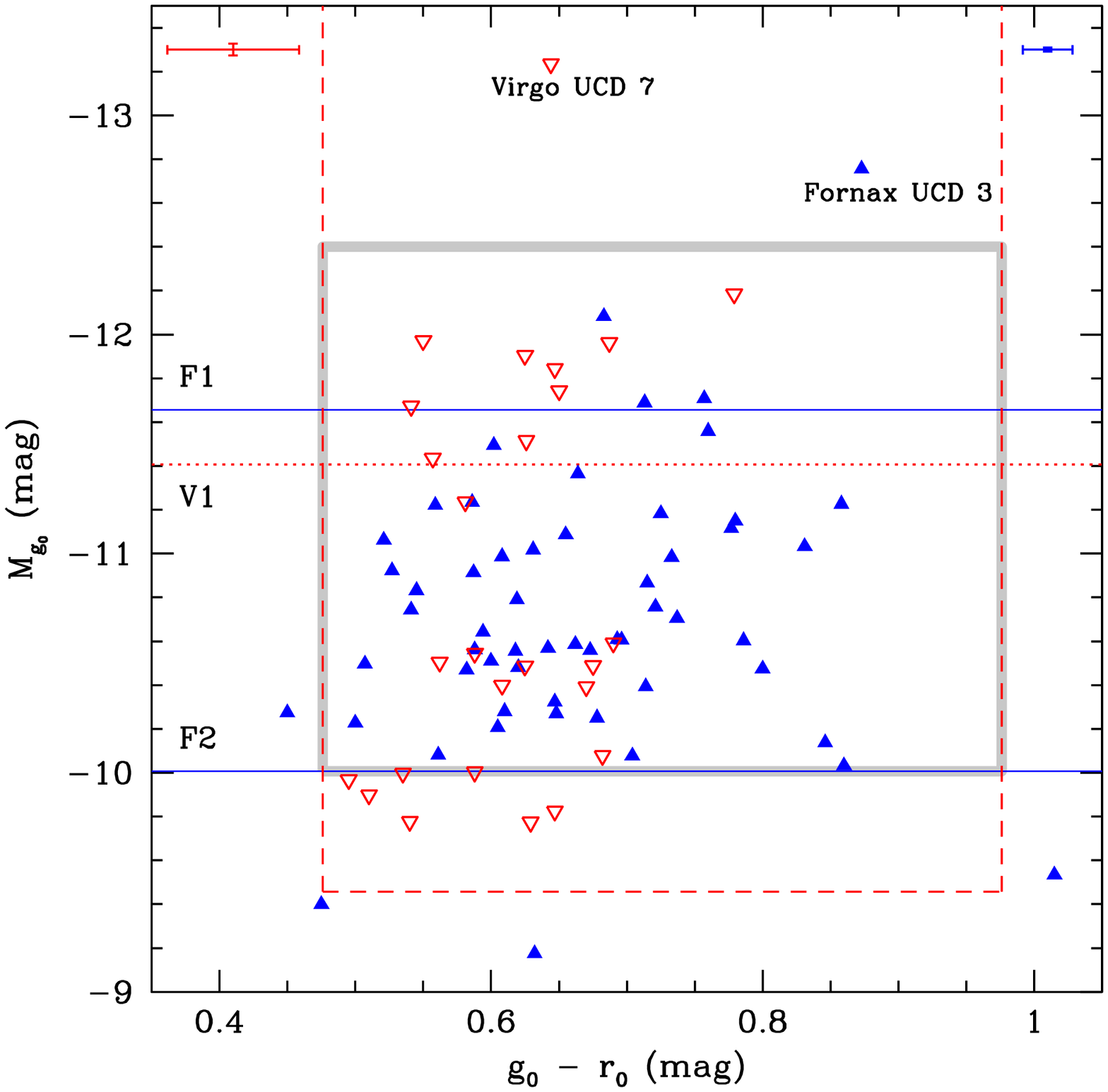}
\caption{Colour-magnitude plot of bright CSSs in the cluster core regions of Virgo (unfilled triangles) and Fornax (solid triangles) from wide-area 2dF and AAOmega observations. Typical photometry error bars are shown at upper-left for Virgo and upper-right for Fornax. The labelled target selection limits are: in Fornax, \citet[F1:][]{Drinkwater..2000a} and our AAOmega NGC 1399 survey with limits the same as \citet[F2:][]{Gregg..2007}; in Virgo, \citet[V1:][]{Jones..2006}. The dashed line shows the extinction-corrected approximate limits of our AAOmega Virgo survey (based on $0.5<(g-r)<1.0$ and $g \le 21.5$). The grey box shows the parameter space common to the 2dF/AAOmega Fornax and Virgo observations. An apparent gap ($-11.2<M_{g,0}<-10.6$) in the luminosity distribution of M87 CSSs is due to low completeness of our AAOmega observations in this magnitude range.}
\label{fig:colmag_2}
\end{figure}

As expected, the extinction corrections \citep*[based on][]{Schlegel..1998} for Fornax were smaller than those for Virgo, due to their very different galactic latitudes. Fig.~\ref{fig:colmag_2} compares extinction-corrected colours and absolute magnitudes of luminous CSSs in the Virgo and Fornax cluster core environments, and shows target selection limits for the various 2dF and AAOmega observations. The luminosity distributions of Virgo and Fornax luminous CSSs in the cluster core region appear to scale with cD galaxy mass (M87 has approximately four times the mass of NGC 1399) -- for example 10 Virgo CSSs are brighter than $M_{g,0}=-11.4$ compared with only 6 Fornax CSSs. To properly compare the luminous CSS distributions in Fornax and Virgo, we define a colour-magnitude range (gray box in Fig.~\ref{fig:colmag_2}) common to the various observation sets; on the colour index axis we use the AAOmega Virgo target selection limits, and on the magnitude axis we use the \citet{Gregg..2007} faint limit and a bright limit of $\mbox{M}_{g,0}=-12.4$ to exclude the unusually luminous CSSs. Our bright limit specifically excludes Fornax UCD3 \citep{Drinkwater..2000a} and Virgo UCD7 \citep{Jones..2006} which are atypically bright for CSSs -- the luminosity of UCD3 appears to be increased by a background spiral galaxy \citep{Evstigneeva..2007}, while UCD7 is bright enough to be a dwarf galaxy.

Colour is a useful measure to compare luminous CSS with GCs in order to investigate how they are related. Studies of extragalactic GC systems \citep[see review by][]{Ashman..1998} show that, although the colour of individual GCs is not a strong function of their luminosity, `broad-band colours reflect the age and metallicity of stellar populations' and the average colours of GC systems trend redder (more metal-rich) with increasing host galaxy luminosity. Our results for M87 and NGC 1399 luminous CSS seem contrary to this trend -- in contrast to the Virgo luminous CSS, Fornax luminous CSS extend to redder colours. The mean $g-r$ colour indices of redshift-confirmed Fornax and Virgo luminous CSSs, within the parameter space common to the 2dF/AAOmega Fornax and Virgo observations (grey box in Fig.~\ref{fig:colmag_2}), are $0.66\pm0.02$ and $0.63\pm0.02$ respectively (with standard deviations of $0.09\pm0.01$ and $0.06\pm0.01$). Although the means appear similar, the Fornax luminous CSS $g-r$ population has a significant red tail not found in the Virgo population -- assuming Gaussian colour distributions, the $t$-test and $F$-test probabilities are only 0.09 and 0.06 respectively that the colour distributions of the luminous CSS populations are indistinguishable; and the non-parametric K-S test probability of 0.10 also confirms this result.

\begin{figure*}
\centering
\includegraphics[width=8.7cm]{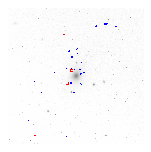}
\includegraphics[width=8.7cm]{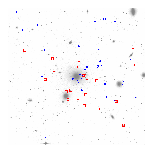}
\caption{Comparison of spatial and radial distributions of Virgo (left: $10^\prime \simeq 50 \; \mbox{kpc}$) and Fornax (right: $10^\prime \simeq 60 \; \mbox{kpc}$) cluster core luminous CSS with respect to their $g-r$ colours. Each DSS blue plate image is 60 arcmin across; North is up and east to the left. The red CSS sub-populations ($g-r \ge 0.7$) are shown as boxes and the blue sub-populations as circles.}
\label{fig:virgofornaxds9}
\end{figure*}

The average colour of individual GCs has been observed to become bluer with increasing radial distance from their host galaxy \citep[see review by][]{Ashman..1998}, although this colour gradient is not observed at larger galactocentric radii. In Fig.~\ref{fig:virgofornaxds9} we compare the wide-area spatial and radial distributions of red ($g-r \ge 0.7$ or $g_0-r_0 \ge 0.68$) and blue Virgo and Fornax luminous CSS sub-populations. By eye it appears that there is a concentration of blue (presumably metal-poor) luminous CSSs near the central massive galaxy in each cluster. However, while the Virgo red sub-population is too small to give reliable statistics, a 2D K-S test confirms (probability statistic 0.1) that the red and blue luminous CSS sub-populations around NGC 1399 have similar overall spatial distributions.

\begin{figure*}
\centering
\includegraphics[width=15cm]{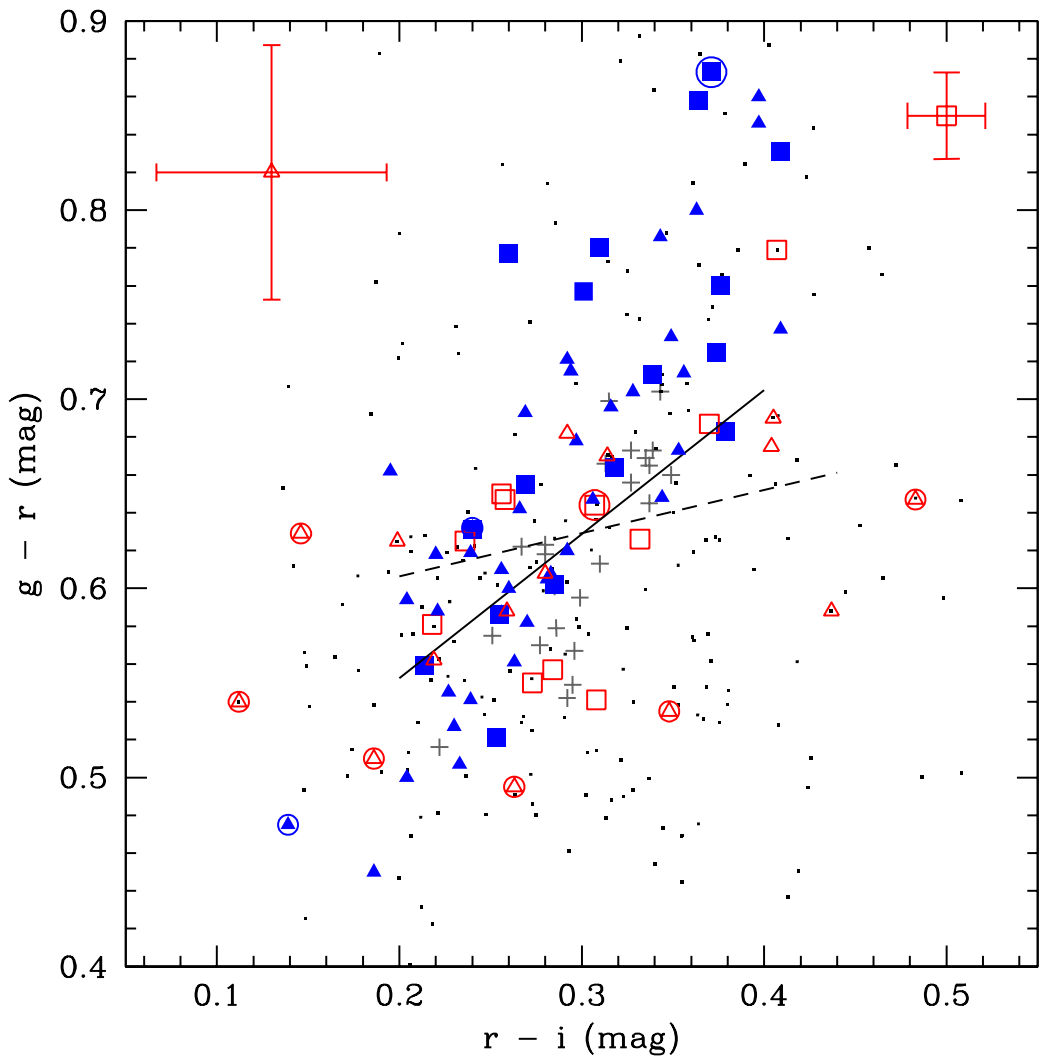}
\caption{Colour-colour diagram of cluster core CSSs in Fornax (filled symbols) and Virgo (open symbols) from SDSS dereddened photometry. The brighter CSSs ($-12.4<M_{g,0}<-11$) are shown as large squares and fainter CSSs ($-11<M_{g,0}<-10$) as smaller triangles. CSSs brighter or fainter than the common Fornax and Virgo observation colour-magnitude limits discussed in the text (see also Fig.~\ref{fig:Virgogri}) are encircled, and are not taken into account in the analysis. For Virgo we show the best fitting lines (continuous and dashed) for the bright and faint CSSs within these common colour-magnitude limits. The distribution of Virgo CSSs are compared with redshift confirmed Virgo cluster core dEs \citep[grey crosses:][]{Binggeli..1985} and M87 GCs \citep[small black squares:][]{Hanes..2001}. Typical 1-$\sigma$ error bars are shown (upper left and right) for the faint and bright Virgo luminous CSS.}
\label{fig:colmag_6}
\end{figure*}

Fig.~\ref{fig:colmag_6} compares the SDSS extinction-corrected colour indices of CSSs in the Fornax and Virgo cluster core environments with those of dEs and GCs. To aid our analysis, we separate the brighter and fainter luminous CSSs at an arbitrary deredenned magnitude of $M_{g,0} = -11$. We also indicate with circles those CSSs outside the magnitude limits (grey box) shown in Fig.~\ref{fig:colmag_2}, and exclude them from our analysis. Fornax luminous CSSs, both brighter and fainter, closely follow the point source (mainly stellar) locus. Virgo luminous CSSs (not circled) are distributed on two best fitting lines -- the brighter CSS's (solid line: $[g-r] = (0.76\pm0.33)[r-i] + (0.40\pm0.10)$) more closely follow the distribution of Virgo dEs, whereas the fainter CSSs (dashed line: $[g-r] = (0.23\pm0.19)[r-i] + (0.56\pm0.06)$) appear more widely dispersed like the M87 GCs. We could interpret the differing distributions of brighter and fainter sub-populations of luminous CSSs in the Virgo Cluster to support the hypothesis that the brightest CSSs come from tidally-stripped dE,N galaxies while the fainter CSSs are the luminous tail of the cluster core GC distribution -- however the dashed line for faint CSS and the GC distribution are influenced strongly by the uncertainties in photometry (see typical error bars on the plot) and small number statistics (evidenced by the uncertainties in best fitting line slopes quoted above).

\subsection{The Intracluster Environment}
IGCs are of interest for several different reasons. By obtaining redshift confirmation of a sample of IGCs, their overall number can be estimated -- if redshift-confirmed IGCs prove to be sufficiently numerous, we can treat them as test particles to map the cluster potential (i.e. the dark matter distribution). Other properties of IGCs can also be compared with intracluster dE galaxies and objects in the cluster core environment to assess whether IGCs are a distinct sub-population with independent origins and kinematics. For example, at large distances from the cluster giant elliptical galaxies, IGC recession velocities should follow those of the overall cluster if they are `freely floating in the cluster potential' as suggested by \citet{West..1995}.

Several CSSs have been spectroscopically confirmed by previous researchers in the Fornax cluster environment closely surrounding the cluster core region. For example, \citet{Bergond..2007} surveyed a $500 \times 150 \; \mbox{kpc}$ area centred on NGC 1399 and found 61 potential IGCs, all except one subsequently shown \citep{Schuberth..2008} to be gravitationally associated with NGC 1399 or its surrounding galaxies. Our AAOmega observations in Virgo and Fornax intracluster fields are the first attempt to survey wide-area ($2^\circ$-diameter) fields for candidate IGCs at least $\sim\!300 \; \mbox{kpc}$ from the cluster centres.

\begin{figure}
\centering
 \includegraphics[width=8.5cm]{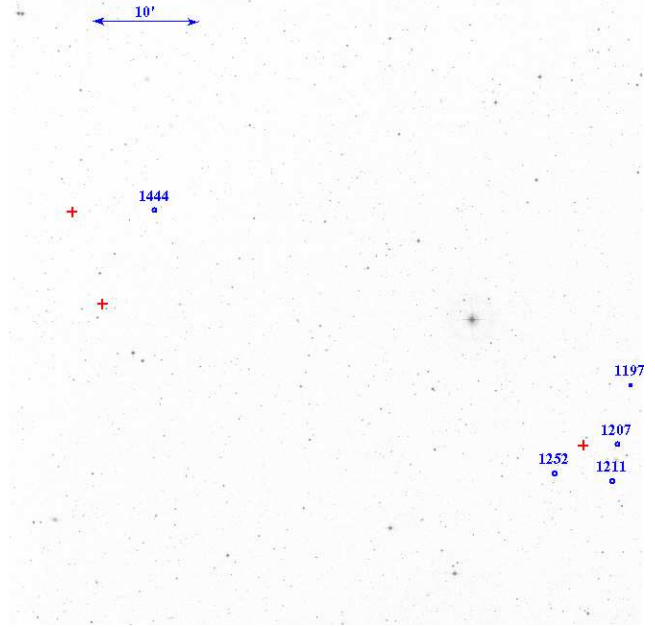}
 \caption{Virgo IGCs and intracluster galaxies. Virgo IGCs (small crosses) are overlaid on a DSS image (POSS-II F) centred at 12:31:30, +09:44:00. North is up and east is to the left. We show potentially associated surrounding galaxies (all are dwarfs if cluster members) from the Virgo Cluster Catalogue \citep[labelled with VCC numbers:][]{Binggeli..1985} that are either confirmed cluster members (VCC 1444) or have unknown recession velocities. The IGC at lower-right is $3 \; \mbox{arcmin}$ ($\sim\!15 \; \mbox{kpc}$) from the nearest dwarf galaxy (VCC 1207, a possible member).}
 \label{fig:aao_1010a}
 \end{figure}

We found three IGCs in the Virgo intracluster field (none of them adjacent to a major galaxy -- see Fig.~\ref{fig:aao_1010a}), but we located no IGCs in the Fornax intracluster field. The Virgo IGCs are sufficiently distant from any major cluster galaxies to be classified as gravitationally unbound. These are the first examples of redshift-confirmed IGCs that can be compared with the population predicted by \citet{West..1995}. Two of the Virgo IGCs have recession velocities of $\sim\!1150 \; \mbox{km} \, \mbox{s}^{-1}$ (similar to that of GCs in the cluster core), and one has an unusually high velocity $\sim\!1980 \; \mbox{km} \, \mbox{s}^{-1}$. Their $g-r$, $r-i$ colours are in the range of CSSs in the Virgo cluster core region (see Fig.~\ref{fig:Virgogri}), although this may be due to our colour selection limits. The IGCs appear as isolated point sources in SDSS images -- as noted in Table \ref{table:aaovirgolist} the brightest one shows hints of resolution on the best SDSS image, with the red, extended source just to the north being very likely a background galaxy.

Table \ref{table:cssintracluster} summarises the results of our AAOmega point source survey of the intracluster environment. We have estimated an upper limit for the population of bright IGCs in the Virgo and Fornax intracluster environments by using the hypergeometric distribution\footnote{The hypergeometric distribution is applicable to a binomial sampling process without replacement. We used the online calculator available at http://www.quantitativeskills.com/sisa/distributions/hypergeo.htm.} with 95 per cent confidence limit. In Fornax we found no IGCs (upper limit 2 IGCs) in observations that were 81 per cent complete to a faint limit of $M_{b_J}<-11.5$, while in Virgo we found 3 IGCs (see Fig.~\ref{fig:aaovirgofornax_3}) to a 54 per cent complete faint limit of $M_{b_J}<-10.2$ suggesting a population upper limit of 9 IGCs.

\begin{table}
\caption{Intracluster/Galaxy Merger CSS Results}
\label{table:cssintracluster}
\centering
\scriptsize{
\begin{tabular}{lcccc}
\hline \hline
$M_{b_J}$ Range & Available & Redshifts & Completeness & CSSs\\
(mag) & Targets & AAOmega & (per cent) & Located \\[3pt]
\hline \hline\\
\multicolumn{5}{l}{Virgo Intracluster Field}\\
 $<-14.0$ & 4 & 1 & 25.0 & - \\
 -13.5 to -14.0 & 97& 43 & 44.3& - \\
 -13.0 to -13.5 & 138 & 55 & 39.9 & - \\
 -12.5 to -13.0 & 146 & 34 & 23.3 & - \\
 -12.0 to -12.5 & 186 & 72 & 38.7 & - \\
 -11.5 to -12.0 & 162 & 123 & 75.9 & - \\
 -11.0 to -11.5 & 185 & 144 & 77.8 & 1 \\
 -10.5 to -11.0 & 277 & 106 & 38.3 & - \\
 -10.0 to -10.5 & 298 & 138 & 46.3 & 2 \\
 -9.5 to -10.0 & 454 & 76 & 16.7 & - \\
 -9.0 to -9.5 & 420 & 10 & 2.0 & - \\
\\
\multicolumn{5}{l}{Fornax Intracluster Field}\\
 $<-14.0$ & 521 & 444 & 85.2 & - \\
 -13.0 to -13.5 & 199 & 148 & 74.4 & - \\
 -12.5 to -13.0 & 247 & 213 & 86.2 & - \\
 -12.0 to -12.5 & 255 & 197 & 77.3 & - \\
 -11.5 to -12.0 & 304 & 228 & 75.0 & - \\
 -11.0 to -11.5 & 350 & 74 & 21.1 & - \\
\\
\multicolumn{5}{l}{NGC 1316 Galaxy Merger Field}\\
 $<-14.0$ & 39 & 31 & 79.5 & - \\
 -13.0 to -13.5 & 13 & 10 & 76.9 & - \\
 -12.5 to -13.0 & 14 & 12 & 85.7 & - \\
 -12.0 to -12.5 & 21 & 9 & 42.9 & - \\
 -11.5 to -12.0 & 26 & 3 & 11.5 & - \\
 \\
\hline
\end{tabular}
}
\end{table}

Our survey of intracluster fields in Fornax and Virgo was limited to the relatively luminous CSSs located at a considerable distance from the cluster centres. We have no theoretical predictions with which to compare these results -- the predictions by \citet{West..1995} relate to the environment of the brightest cluster galaxy. Our low IGC count may be due to the areal limit of our survey, since outside the cluster core the distribution of IGCs may be non-uniform, mirroring the cluster galaxy distribution. Our findings suggest that populations of IGCs may exist in both clusters, but that we need to probe a wider area of the intracluster regions to a deeper faint limit in order to further investigate their distribution and origins.

\subsection{The Galaxy Merger Environment}
\label{galaxymerger}
Active gas-rich galaxy mergers are observed to contain close groupings of newly-condensed star clusters \citep[such as those observed in the Antennae galaxy pair][]{Kroupa..1998}, and numerical simulations \citep[e.g.][]{Fellhauer..2002} confirm that the largest of these should survive disruptive tidal forces and merge on relatively short time-scales ($\sim\!100 \; \mbox{Myr}$) to form stellar superclusters. Superclusters with mass of order $10^8 \; \mbox{M}_{\odot}$ may then evolve into present-day bright CSSs, or potentially into spheroidal dwarf galaxies with low dark matter content. This mechanism for luminous CSS formation is theoretically supported by simulations \citep{Fellhauer..2002, Fellhauer..2005}, whereas their formation as ancient but unusually massive GCs is not well understood. On the other hand, it is unclear whether stellar superclusters are ejected with enough efficiency from their formation sites to account for the CSS populations we find in cluster core and intracluster environments.

The NGC 1316 galaxy merger environment borders our Fornax intracluster field (see Fig.~\ref{fig:aaovirgofornax_11}). \citet{Goudfrooij..2001a, Goudfrooij..2001b} found evidence of both old and intermediate age (high hydrogen content) GCs within an $8^\prime\times8^\prime$ region surrounding this 3-Gyr-old merger remnant galaxy, including four exceptionally bright CSSs ($M_V<-12$). Although we radially restrict our definition of the NGC 1316 galaxy merger environment to $<\!20 \; \mbox{arcmin}$ ($130 \; \mbox{kpc}$, encompassing the faintest traces of the galaxy's stellar envelope), our AAOmega intracluster field extends east of NGC 1316 almost $2^\circ$ ($800 \; \mbox{kpc}$). Above the relatively bright limit ($M_{b_J}<-11.7$) of our survey, we found no evidence of dispersed intermediate-age stellar superclusters emanating from the NGC 1316 galaxy merger. While observing conditions restricted our detection limit to the most luminous CSSs, we would have expected to find any ejected stellar superclusters since their initial luminosity ($M_B\simeq-15$) should still be above our detection limit after $3 \; \mbox{Gyr}$ \citep{Fellhauer..2002}. We conclude that there is no evidence such objects have been dispersed outside the innermost region of the NGC 1316 galaxy merger or into intracluster space.

The galaxy NGC 4438, which has been strongly distorted by interaction with NGC 4435, is marked in Fig.~\ref{fig:aaovirgofornax_11} at the edge of our M87 field. In our survey we found no overdensity of confirmed foreground or cluster objects that might be associated with these galaxies.

\section{Summary of Findings}
In our search for CSS, we have obtained redshift measurements of colour-selected point sources in two nearby galaxy clusters. Our new observations:

\begin{itemize}
\item Extend the faint limit of previous CSS redshift surveys in the Virgo cluster core by approximately one magnitude to $M_{g,0} \le -9.7$.
\item Survey both Virgo and Fornax wide-area intracluster fields for the first time.
\item Extend the search for CSSs evolved from stellar superclusters to an increased radial distance from the galaxy merger remnant NGC 1316.
\end{itemize}

Amalgamating our results with previous CSS data, our key findings are as follows:

\begin{itemize}
\item We have significantly increased the number of spectroscopically-confirmed luminous CSSs in the Virgo Cluster core region, enabling a more informed comparison with luminous CSSs in the Fornax Cluster core. We found only two additional luminous CSS in the Fornax Cluster core region, but improved the accuracy of recession velocity measurements for five previously located CSS.
\item The estimated total population of luminous CSS ($M_{b_J} \le -10.5$) in the spatially equivalent core region of M87 is approximately half that of NGC 1399, despite M87's substantially greater mass and GC specific frequency. As an isolated result, this implies that either the process of production of such objects in NGC 1399 is more efficient, or the process of destruction in M87 is more efficient.
\item We have confirmed that the velocity dispersion of luminous CSSs in both clusters is significantly less than that of either the cD galaxy GC system or the cluster dwarf galaxies.
\item Using extinction-corrected $gri$ photometry, we find that Fornax has a sub-population of luminous CSSs that are redder (presumably more metal-rich) than those found in Virgo. The red and blue Fornax luminous CSS have similar spatial and radial distributions around NGC 1399. Fornax CSSs and the brightest Virgo CSSs follow the dE distribution in colour-colour space, whereas fainter Virgo CSSs have a more dispersed colour-space distribution similar to the M87 GCs.
\item We found no bright IGCs in Fornax but uncovered three IGCs in Virgo -- these are the first spectroscopically confirmed IGCs that are not associated with the cluster core or any neighbouring non-dwarf cluster galaxy. Allowing for the different faint survey limits of our Virgo and Fornax intracluster observations, IGCs may exist in both the Virgo and Fornax clusters. A multi-fibre spectroscopy survey with a larger aperture telescope will be required to locate and properly compare IGC populations in both clusters to an adequate faint limit.
\item Whereas previous researchers have found several luminous CSSs within 4 or $5 \; \mbox{arcmin}$ of NGC 1316, we found no additional luminous CSSs in a radial arc extending between approximately 10 and $20 \; \mbox{arcmin}$ to the east of NGC 1316 or in the adjacent intracluster field. Although the aereal coverage of our field around NGC 1316 was limited, our result suggests that if intermediate age luminous CSSs are formed from the merging of several smaller CSSs created in NGC 1316's galaxy merger, they have not been widely dispersed.
\end{itemize}

\section*{Acknowledgments}
We thank Rob Sharp, Will Saunders, Heath Jones and the other Anglo-Australian Telescope staff who assisted our AAOmega observations, and J. Bryn Jones (University of London) for the ideas behind Fig.~\ref{fig:virgofields}. We especially thank the reviewer G. Bergond for his constructive comments and suggestions.

Part of the work reported here was done at the Institute of Geophysics and Planetary Physics, under the auspices of the U.S. Department of Energy by Lawrence Livermore National Laboratory in part under Contract W-7405-Eng-48 and in part under Contract DE-AC52-07NA27344.

\bibliography{AAO2006_mnras}

\bsp

\label{lastpage}

\end{document}